\documentclass[twocolumn]{aastex631}
\usepackage{graphicx}
\usepackage{amsmath}
\usepackage{verbatim}
\usepackage{subfigure}
\usepackage{float}

\received{February 28, 2024}
\accepted{September 12, 2024}

\begin{document}

\title{Studying The Effect of Radiation Pressure on Evolution of a Population III Stellar Cluster}

\author{Sukalpa Kundu}
\email{sukalpa.k123@gmail.com}
\correspondingauthor{Sukalpa Kundu}
\author{Jayanta Dutta}
\affiliation{Harish Chandra Research Institute, Chhatnag Rd, Jhusi, Prayagraj, Uttar Pradesh 211019}
\begin{abstract}

Recent numerical simulations have shown that the unstable disk within the central regime of the primordial gas cloud fragments to form multiple protostars on several scales. Their evolution depends on the mass accretion phenomenon, interaction with the surrounding medium and radiative feedback respectively. In this work, we use a fast semi-analytical framework in order to model multiple protostars within a rotating cloud, where the mass accretion is estimated via a Bondi-Hoyle flow and the feedback process is approximated through radiation pressure. We observe that while some of the evolving protostars possibly grow massive ($\approx 1-75\ M_{\odot}$) via accretion and mergers, a fraction of them ($\approx 20\%$) are likely to be ejected from the parent cloud with a mass corresponding to $M_{*} \lesssim 0.8\ M_{\odot}$. These low-mass protostars may be considered as the potential candidates to enter the zero-age-main-sequence (ZAMS) phase and possibly survive till the present epoch.
\end{abstract}

\section{Introduction} 
\label{sec:intro}

In the standard framework of cosmology, primordial fluctuations in the density field of the universe grow into dark matter (DM) minihalos (of virial mass $\approx10^5-10^6\ M_{\odot}$) as a consequence of the hierarchical structure formation \citep{Navarro+96,Haiman_2011,Wise_19,Wang+20,Springel+21}. Baryonic matter including the electrons from the recombination era settles into the potential wells of these minihalos to form the primordial gas clouds, which are the sites of the Population III (or Pop III) stars \citep[see e.g., the reviews][]{Barkana_Loeb_01,Bromm+04,Ciardi_Ferrara_05,Klessen:2023qmc}. Subsequently, this unstable gas undergoes nonlinear collapse as a result of self-gravity and thermodynamical instabilities through the primordial chemical network \citep{Palla_1983,Glover_Abel_08,Turk_2011,Bovino+14,Dutta_2015,Barkana+18}, eventually giving rise to a disk-like structure that fragments to form multiple protostars \citep{Stacy+10,Clark+11,Greif_2011,Dutta_15_3,Sharda+19,Inoue+20,Wollenberg_20,Chiaki+22}. At some epoch, the mass accretion process of these protostars is affected by the radiative feedback that becomes a crucial factor influencing their mass evolution \citep{Hosokawa+11,Stacy+12,Hirano+15,Latif_2022,Sugimura+23}. 

Depending on the initial configuration such as rotation, thermal and chemical instabilities, some of these protostars continue to grow massive either through accretion \citep{Haemmerle+17,Woods_17} or through mergers \citep{Kulkarni+19,Susa19}. Some of these stars can go on to explode as pair-instability-supernovae (PISN) \citep{Chen_2014,Yoshii_2022,Padmanabhan+22,Venditti+24} or collapse to form the seed of the blackhole \citep{Smith_2018,Safarzadeh+2020,Santoliquido_23} or undergo runaway merger with other stars \citep{Vergara21,Seguel+2020}. A fraction of the fragments can also lead to the formation of low-mass stars that may even escape the potential well of the star cluster \citep{Marigo+01,Komiya+15,Ishiyama_2016,Dutta16,Raghuvanshi+23}.

 However, to investigate the initial mass function (IMF), we need to follow the evolution for hundreds of thousands of years, which seems to be non-trivial in existing 3D numerical simulations. This is mainly due to complexities such as complicated chemical networks and the high dynamic range in densities. Introducing the radiative transfer to account for the feedback channels such as radiation pressure, photoionization and photodissociation also adds to the computational expenses \citep[see review by][]{Haemmerle+20}. Besides, most of the modern simulations do approach the very high densities ($10^{14}\rm \ cm^{-3}$) as the numerical resolution for the protostellar density \citep{Jaura+18,Latif_2022}. Although these studies evolved for several kiloyears, there is still no consensus on the role of radiative feedback in regulating accretion \citep{Jaura+22}. There are though a few studies \citep{Greif+12,Becerra_15,Prole_21} which consider an extremely high-density regime ($10^{19}\rm \ cm^{-3}$) as the protostellar density. While the extremely-high resolution simulations could be evolved only for a few hundred years, it was only the relatively low-resolution simulations ($\lesssim10^{13}\rm \ cm^{-3}$) that could be run for a few kiloyears \citep{Susa_13,Stacy+16,Latif_2022}. Another issue with the high-resolution simulations is to achieve a very narrow parameter space exploration. However, in general, the star formation process is inherently stochastic, which requires repeated calculation incorporating this stochasticity, i.e., with different turbulent realizations, for instance. As a result, we still do not have a complete understanding of the initial mass function of the Pop III stars.

In that regard, the purpose of our work is to develop an alternative numerical framework that can capture the entire accretion, feedback, merger and ejection phenomenon, respectively. In an earlier work, \citet{Dutta+2020} proposed a semi-analytical model using Bondi-Hoyle flow in order to investigate the long-term interaction of a single protostar with the ambient gas medium. Although that work explored a pathway towards the possibility of the existence of low-mass stars till the present epoch, it could not address the final mass of the protostars due to the absence of the radiative feedback. As a result, the continuous mass accretion resulted in runaway instabilities. In this work, we attempt to improve the previous model by including (i) rotation in the gas cloud, (ii) a multi-particle system, (iii) radiation pressure as the feedback channel, (iv) a dynamic model for protostellar radius instead of a point mass and (v) protostellar mergers, respectively.

In the next section \S \ref{sec:numerical_methodology}, we describe in detail the numerical methodology that comprises the full set of equations governing the time evolution of the N-Body system. The dynamical evolution of all the relevant quantities including the formation of binary stars is outlined in section \S\ref{sec:results}, along with a brief overview of the stability of our model in section \S\ref{sec:stability}. A comprehensive review of the statistical trend emerging from our model is presented in section \S\ref{sec:stat}. We summarize our results in section \S\ref{sec:discussions}, followed by a detailed discussion and observational evidence that signify our work. Additionally, we present a few numerical tests that we performed to ensure the correctness of our simulations in the Appendix \S\ref{sec:test}. The dynamical evolution of the protostars described in section \S\ref{sec:stat} are further illustrated in Appendix \S\ref{sec:diff_ic}.

\section{Numerical Methodology} 
\label{sec:numerical_methodology}

In this section, we describe the numerical set-up that allows us to study the long-term evolution of the system, depicted as a supersonic, compressible flow coupled with gravity.

\subsection{Rotation of the gas Cloud} 
\label{sub:cloud}

Following the prescription in the Larson-Peterson similarity solution in the case of a polytropic equation of state, i.e. the density gradient can be expressed as $\partial \rho/\partial r = -2/(2-\gamma)$, where the polytropic index $\gamma\approx1.1$ \citep{Larson+69,Omukai+98}. Considering the gas cloud of size $2.7$ pc, the density profile can numerically be modeled as $\rho(r) \approx \rho_0 /(1+r/r_0)^{2.2}$, where $\rho_0 \approx n_0 \mu m_{\rm p}$, $n_0 \approx 10^{13}\ \rm cm^{-3}$, $\mu \approx 2.33$, $r_0 \approx 5\, \rm AU$, and $m_{\rm p}$ being the proton mass. In the same token, the temperature and the local sound speed can be modelled as $T(r)\approx1200 (\rho(r)/\rho_0)^{\gamma-1}\ K$ and $c_s(r) \approx \left( 5 K_B T(r) / 3 \mu m_{\rm p} \right)^{0.5}$ \citep{Abel:2001pr,Yoshida_08,Dutta+2020}.

It is to be noted that the dependence of the density and the temperature profile is motivated by the thermodynamics of the spherical collapse problem through the primordial chemical network. Initially, the temperature remains at $200$ K, as the gas is in local thermodynamic equilibrium (LTE) with the molecular fraction $\approx10^{-3}$. However, as the gas collapse proceeds to high density ($10^8 \rm \ cm^{-3}$), the atomic hydrogen converts to fully molecular as a result of the three-body reaction through the primordial chemical network, heralding the temperature $1200$ K (Fig. 3, \citet{Yoshida_2006}).

\begin{figure}
\centering
\includegraphics[width=0.45\textwidth]{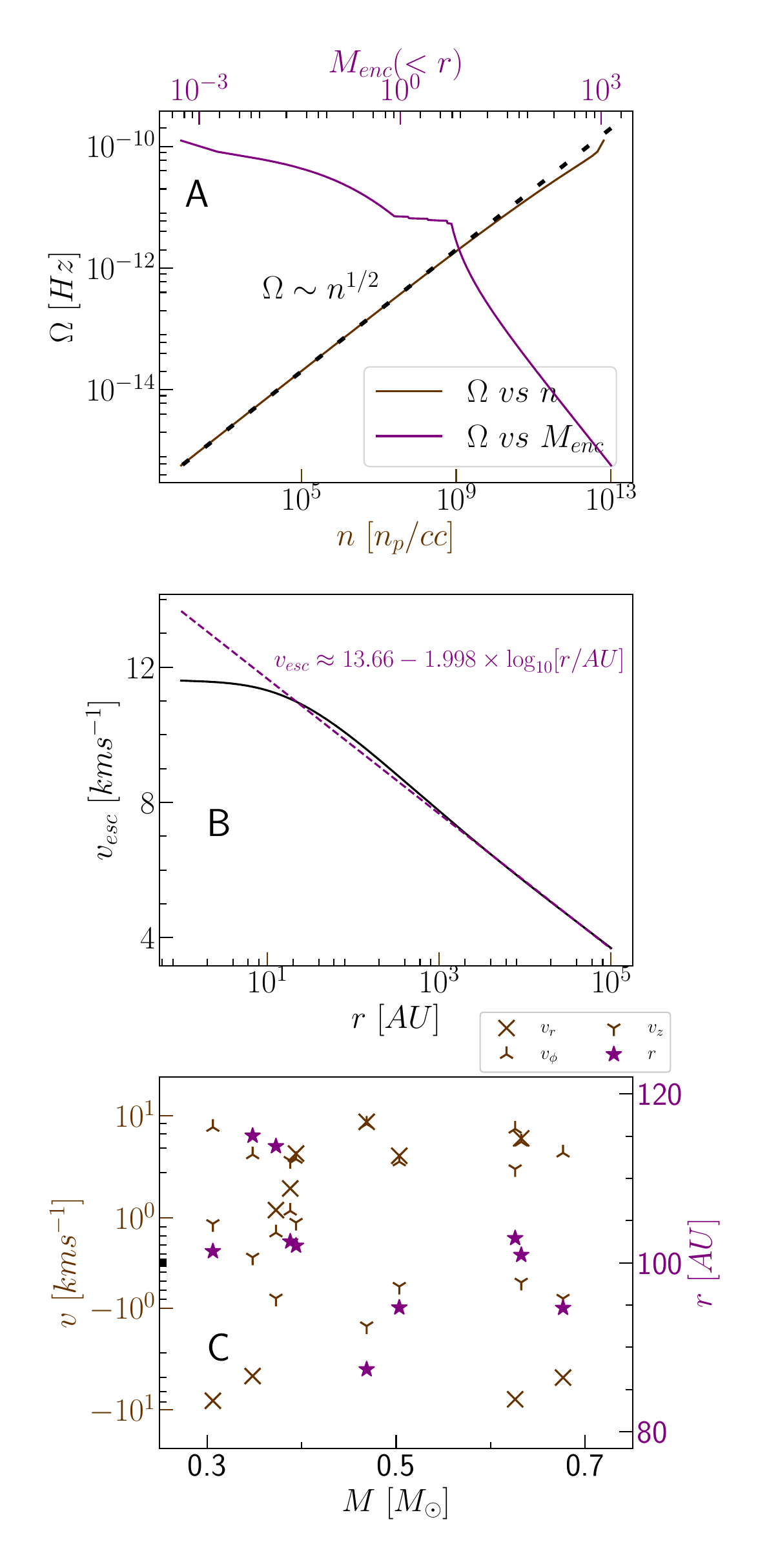}
\caption{Top panel demonstrates the variation of the angular velocity with the number density of the cloud, shown by the brown line. We fit this variation (black dashed line) with a power-law characteristic ($\Omega \sim n^{0.5}$). The purple line represents the variation of the same as a function of the total enclosed mass that comprises both gas and protostars. In the middle panel, the escape velocity ($v_{\mathrm{esc}}$) has been numerically calculated (represented by the black line), which corresponds to the logarithmic fit ($v_{\mathrm{esc}}\approx13.66-1.998\log[r/\mathrm{AU}]$) denoted by the purple line. We observe a deviation within the very central regime. Hence, the fit corresponds to the escape velocity at a radial distance $r\gtrsim30$ AU. The bottom panel shows the distribution of the initial radial distance (purple markers) and three components of the velocity $v_r,v_\phi,v_z$ (brown markers) corresponding to the initial mass of the protostars.}
\label{fig:IC}
\end{figure}

We set the cloud to be rotating around the $z$-axis with the rotational speed given by $v_{\rm rot} \approx g_{\rm rot} \left( G M_{\rm enc}(r) /r\right)^{0.5} (\sqrt{r^2-z^2}/r)$, where $g_{\rm rot}\approx 0.2$ is defined as the rotation parameter. This factor implies that the gas cloud does not rotate with full Keplerian speed \citep{Yoshida_2006}. The strength of rotation can be determined by the ratio of rotational to gravitational energy, estimated as $\beta \approx 0.005$ \citep{Sterzik+03,Machida+08}, which implies a slow-rotating system. Figure \ref{fig:IC}A shows the variation of the angular velocity of the cloud as a function of the density \citep[$\Omega \sim n^{0.5}$, in accordance with earlier works by][]{Matsumoto_1997,Dutta:2015tcl}. Additionally, the plot illustrates the correlation between the angular velocity and the total enclosed mass ($M_{\rm enc}(<r)$), which includes both gas and protostars. In the next section, we describe the dynamics of the protostars. 

\subsection{Modelling the dynamics}
\label{subsec:dynamics}

Our system can be considered as a set of multiple gravitating protostars of mass $M_i$, position $\vec{x}_i$ and velocity $\vec{v}_i$, accreting the ambient gas at a Bondi-Hoyle accretion rate \citep{EDGAR04}. As the protostars roam around the cloud, their motion is influenced primarily by three sources, (i) the gravitational force exerted by the gas, (ii) the N-Body interaction with their counterparts and (iii) the dynamical friction arising from the acquired momentum through the accreting mass. Accounting for all these processes, we have solved the following set of equations for the N protostars to obtain the dynamics.

\begin{align}
\frac{dM_{i}}{dt} &= \frac{4 \pi G^2 M_{i}^2 \rho(r)}{[V_{\rm i,eff}]^3}
\label{eq:acc}
\end{align}

\begin{align}
\frac{d^2 \vec{x}_i}{dt^2} &= -\frac{G M_{\rm enc}(r_i)}{r_i^2} \frac{\vec{x}_i}{r_i} - \frac{(\vec{v}_{i} - \vec{v}_{i,\text{gas}})}{M_{i}} \frac{dM_{i}}{dt} + \vec{F}_{i,\mathrm{int}},
\label{eq:dyn}
\end{align}
where $\vec{v}_i=d\vec{x}_i/dt$ and $r_i=|\vec{x}_i|$, denoting the velocity and radial distance of the i'th protostar from the centre of the cloud in the gravitational potential well of the entire system.

Eq. \ref{eq:acc} represents the Bondi-Hoyle accretion rate where the $V_{i,\mathrm{eff}}$ implies the effective velocity of the protostar considering their $x,y,z$ velocity components along with the sound speed and rotational velocity of the gas, defined in Eq. \ref{eq:veff}. In the Eq. \ref{eq:dyn}, the first term denotes the cloud's gravity, where $M_{\mathrm{enc}}(r_i)$ signifies the total enclosed gas mass at a distance $r_i$ from the centre of the cloud, defined in Eq. \ref{eq:menc}. The second term is for dynamical friction where $\vec{v}_i-\vec{v}_{gas}$ implies the velocity of the protostar relative to the gas. The third term denotes the interaction of the $i$'th protostar with the other N-1 protostars, given by Eq. \ref{eq:fnbody}.

\begin{equation}
M_{\text{enc}}(r) = \int_0^r dx \rho(x) 4\pi x^2
\label{eq:menc}
\end{equation}

\begin{equation}
[V_{\text{i,eff}}]^2 = c_s^2 + (\vec{v}_i - \vec{v}_{i,\text{gas}})^2
\label{eq:veff}
\end{equation}

\begin{equation}
\vec{F}_{i,\text{int}} = - \sum_j G m_j (\vec{x}_i - \vec{x}_j)/r_{ij}^3
\label{eq:fnbody}
\end{equation}

\begin{equation}
r_{ij}^2 = (\vec{x}_i - \vec{x}_j)^2
\label{eq:rij}
\end{equation}

The rotational velocity of the cloud $v_{\mathrm{rot}}$ in spherical polar coordinate can be represented in the cartesian system as $\vec{v}_{i,\mathrm{gas}}$, where $i$ represent the position of the protostars. This transformation is done using the following equation:
\newline
\begin{equation}
\vec{v}_{\mathrm{i,gas}}=[v_{\mathrm{rot}} (-\frac{y_i}{\sqrt{x_i^2+y_i^2}}),v_{\mathrm{rot}} (\frac{x_i}{\sqrt{x_i^2+y_i^2}}),0], 
\end{equation}

where the three terms represent the $x,y\ \&\ z$ components of the gas velocity, respectively.

We have tested the accuracy of our dynamical evolution model (for details please see Fig. \ref{fig:acc} in the Appendix \S\ref{sec:test}). The N-Body interaction has also been tested rigorously with the well-known `Figure-8 model' for the three-body problem \citep{Chenciner2000}, which is shown in Fig. \ref{fig:3b} in the Appendix \S\ref{sec:test}. In the next section \S\ref{sub:dt}, we describe the algorithm we have used to integrate the above set of equations.
\subsection{Adaptive timestep}\label{sub:dt}

Due to the nonlinear behaviour of this dynamical system, the integration of the equations becomes highly sensitive to the choice of the time step. As the protostars interact among themselves and also with the ambient medium, the solutions to the above set of equations require an extremely variable (i.e., adaptive) time step which determines the stability of the system. For example, tiny timesteps are crucial in the case of close encounters when the motion changes rapidly. Besides, an infinitesimal timestep is also required in order to accurately capture the merging phenomenon.

We implement a fourth-order Runge-Kutta algorithm with adaptive time-stepping criteria based on individual protostellar velocities and accelerations. The choice of this time-stepping scheme satisfies the well-known Courant–Friedrichs–Lewy (CFL) criteria. First, we calculate the timestep for each protostar and use the smallest of them at each step. We calculate the individual time steps as

\begin{equation}
\Delta t_i=\epsilon\ \left(\frac{v_{x,i}^2+v_{y,i}^2}{\left(\frac{dv_{x,i}}{dt}\right)^{2}+\left(\frac{dv_{y,i}}{dt}\right)^{2}}\right)^{\frac{1}{2}}.
\label{timestep}
\end{equation}

where $\epsilon = 2 \times 10^{-3}$ \citep[see e.g.,][]{Dutta+2020}. Finally, we choose the minimum of these timesteps for each star as the global timestep. Mathematically, it is represented as $dt \equiv min(\Delta t)$, which ensures the stability of the system.

\subsection{Escape velocity}
\label{sub:escape}

As we integrate the system with the above time-stepping scheme, a few protostars may move away to the outer periphery of the cloud and even get ejected. Here, we demonstrate the variation of the escape velocity across the gas cloud.

Assuming $U(r)$ as the potential energy at a distance $r$ from the centre of the cloud, we intend to estimate the escape velocity using this formula $v_{\mathrm{esc}}=\sqrt{-2U(r)}$, where $U(r)$ is defined as 

\begin{equation}
U(r)=-\int_{\infty}^{r} dx GM_{\mathrm{enc}}(x)/x^2.
\end{equation}

Fig. \ref{fig:IC}B shows the variation of the escape velocity as a function of the radial distance from the centre. In general, a protostar having a velocity exceeding the escape velocity ($v>v_{\mathrm{esc}}$) is likely to be ejected from the gas cloud. However, given the possibility of the formation of binary pairs (see Sec. \ref{sub:binary}) that may revolve around each other with very high velocities, $v>v_{\mathrm{esc}}$ may not be the sufficient criteria to mark the escaping protostars. Hence, in our simulation, we mark a protostar as escaped when its radial distance from the centre of the cloud exceeds $r>2.7$ pc, i.e., the size of the cloud.

Unless we have a feedback mechanism that halts the mass accretion, the protostars keep on accreting and their masses tend to increase indefinitely. In the next section, we describe the feedback model that we have used in our work.

\subsection{Radiation Pressure}
\label{sub:feedback}

In our work, we have used radiation pressure as the channel for the feedback. To estimate its effects, we first need to calculate the amount of energy emitted from the accreting protostar, which is given by the accretion luminosity $L=f (GM\dot{M}/R)$ \citep[see, e.g.,][]{Smith+11,Stacy+16} \footnote{It is to be noted that in reality, luminosity of any star is estimated from the amount of radiation coming from its interior which could be very different from the accretion luminosity and possibly dominate for more massive stars \citep[see e.g.,][]{Haemmerle+18}.}. Here R denotes the radius of the protostar of mass M accreting at a rate $\dot{M}$. The factor $f$, which determines the fraction of energy being radiated, has a value of $f\approx1$ for a spherical accretion model and approaches $f\approx0.1$ for non-spherical disk accretion \citep{Stacy+16}. We have used a spherical accretion case in our work through the parameter $f\approx1$. This is primarily because radiation pressure can halt the mass inflow on the protostar only in the case of spherical accretion \citep{Kuiper+10,Wolfire+87}.
 
To effectively model the radiation pressure, one needs to address a specific issue. As the radiation from the protostar carries a momentum, a fraction of it is likely to be absorbed by the surrounding medium depending on the opacity $\chi$ \citep{Rybicki+86}. Consequently, the gas in the nearby regime experiences a net force in the radially outward direction. Once this outward force tends to supersede the self-gravity of the protostar, the mass accretion process becomes substantially affected in the case of spherical accretion. In our semi-analytical model, we capture this complicated phenomenon by introducing the well-established Eddington limit, which sets the maximum rate at which a protostar can accrete the ambient medium. Mathematically, this condition is achieved at an accretion rate of

\begin{equation}
\dot{M}_{\mathrm{Edd}}\approx \left(4 \pi R c/ f k_0 \chi_e \right).
\end{equation}

Here $\dot{M}_{\mathrm{Edd}}$ symbolises the Eddington limit of accretion and $R,c$ are the protostellar radius and the speed of light, respectively \citep{Omukai_2003}. The term $\chi_e\approx0.398 \ \mathrm{cm^2/g}$ denotes the Thompson opacity, or the free-free opacity, and we have taken the specific opacity $k_0\approx10$ \citep{Krumholz+09,Kuiper+10,ARC10} \footnote{Despite the feedback process to be manifested here is the radiation pressure at the photospheric surface of the protostar, there are several other feedback channels used in a number of studies \citep[see e.g.,][]{Klessen:2023qmc}. For instance, the UV radiation from the massive protostars can photo-evaporate the gas and hence is likely to provide the radiation pressure on the gas cloud. Thus it can hinder the growth of the star by removing the gaseous fuel \citep{Jaura+18}. Although this is the mechanism on which the numerical simulations mostly focus, this is difficult to incorporate in a semi-analytical model.}.

Once the protostar reaches the Eddington limit in our model, the key question to address is at which value we should suppress the accretion rate \citep{Wolfire+87,Kuiper+10}.

For example, the accretion rate is considerably low $\approx 10^{-9}-10^{-10} \ M_{\odot}yr^{-1}$ in the case of escaping protostars, but is as high as of the order of $10^{-1}-10^{-2}\ M_{\odot}yr^{-1}$ for the protostars roaming around the central regime. So, which value to take to simulate the accretion rate in the post-Eddington limit? Numerically, this can be achieved by choosing an intermediate value $\dot{M}_{\mathrm{sup}}\approx10^{-5}\ M_{\odot}yr^{-1}$ that corresponds to a suppressed accretion rate once it exceeds the Eddington limit. The bottom panel of Fig. \ref{fig:vary} details the sensitivity of the results on the choice of this suppressed accretion rate. Next, we discuss the model for the protostellar radius, necessary to understand the feedback process.

\subsection{Protostellar radius}
\label{sub:star_radius}

In this work, we have improved the previous work by \citet{Dutta+2020} by considering a model for the protostellar radius that can be thought as a sink particle \citep{Bate+95} used in 3D simulations. As the protostars in our model are still in the evolving phase with ongoing accretion, their radii are expected to vary continuously depending on the mass and accretion rate. We have obtained the estimation of the radius following the prescription used in \citet{Stacy+12}. Below we describe this in detail.

In the initial adiabatic accretion stage, the radius is approximated with 

\begin{equation}
R_I\approx 50 R_{\odot} ( M_{*}/M_{\odot})^{1/3} ( \dot{M}_{*}/\dot{M}_{\mathrm{fid}} )^{1/3},
\end{equation}

where $\dot{M}_{\mathrm{fid}}=4.4 \times 10^{-3}\ M_{\odot}yr^{-1}$ is a fiducial accretion rate used in the simulation by \citet{Stacy+12}. In this stage, the protostellar radius keeps on increasing depending on the mass and accretion rate. 

In the later stages, when the protostar becomes massive, it enters the process of Kelvin-Helmholtz (KH) contraction. At this stage, the radius is given by

\begin{equation}
R_{II} \approx 140 R_{\odot} ( M_{*}/10 M_{\odot} )^{-2} ( \dot{M}_{*}/\dot{M}_{\mathrm{fid}} ).
\end{equation}
We set the radius to $R_{II}$ when it falls below $R_{I}$. Finally, this contraction comes to an end when the protostar reaches the Zero-Age-Main-Sequence (ZAMS) stage. At this stage, we set the radius to $R_{\mathrm{ZAMS}}$, defined below in equation \ref{eq:rzams}.
  
\begin{equation}  
R_{\mathrm{ZAMS}} \approx 3.9 R_{\odot} (  M_{*}/10 M_{\odot} )^{0.55}. 
\label{eq:rzams}
\end{equation}

In our simulation, we set the radius to be of the order
\newline
$R=max[min(R_{I},R_{II}),R_{ZAMS}]$

\subsection{Merger}\label{sub:merger}

Estimating the radius of the protostars allows us to study the merging phenomena between them. We have modelled the merger with a perfect inelastic collision following \citet{CHAMBERS98}. We consider two protostars with mass $m_i,m_j$, position $\vec{x}_i,\vec{x}_j$, velocity $\vec{v}_i, \vec{v}_j$ and radii $R_i,R_j$. When the distance between these protostars, denoted by the notation $r_{ij}\approx|\vec{x}_i-\vec{x}_j|$, falls below a distance that is the sum of their radii $R_{i}+R_{j}$, we merge them to form a single protostar.  The merged one has the following properties:
\newline
\newline$M=m_i+m_j$, 
\newline$\vec{v}=(m_i \vec{v}_i + m_j \vec{v}_j)/(m_i+m_j)$, and 
\newline$\vec{x}=(m_i \vec{x}_i + m_j \vec{x}_j)/(m_i+m_j)$.
\newline
\newline
We model the merger in such a way that the other one is removed from the system. Besides, we have tested the implementation with a simple inelastic collision, described in the Appendix \S\ref{sec:test}, in Fig. \ref{fig:2b}. It is to be noted that the merging model we used here is actually an oversimplification of the actual physical process of the stellar mergers. For example, the tidal radius can also be used to determine the merging phenomenon. There may be other criteria and additional conditions, which could have a significant impact on how much merging was to occur, thus determining the merger rate \citep[see e.g.,][]{Wollenberg_20, Li+20}.

Here, we would like to emphasize that in traditional hydrodynamical simulations, merger happens between the two sink particles mimicking the protostars, whose size is of the order of AU \citep{Bate+03}. Our model is able to capture the merger even at a scale of $R_{\odot}$, which is much more accurate to address the problem. This has become possible only in a semi-analytical calculation, of course with few approximations. Finally, we are in a position to describe the initial distribution of the protostars.

\subsection{Initial Conditions}
\label{sub:ic}

We initialize our system with a total of 10 protostars. Figure \ref{fig:IC}C illustrates the distribution of the mass, position, and the three components of velocity. Here we briefly mention the motivation behind choosing our initial conditions below.

Once the gas density reaches the protostellar density ($\approx10^{13}\rm \ cm^{-3}$, set by the numerical resolution) and the corresponding temperature reaches around 1200K, it is separated from the rest of the gas with a mass $M_{J}(n,T)\approx0.05\ M_{\odot}$, characterized by the Jeans criteria. This means any new fragment will have an initial mass of the order $10^{-2}\ M_{\odot}$ that has the velocity components in the range $\approx0-10$ km/s within a regime of $\approx100$ AU from the centre, depending on the initial configuration of the cloud \citep[see e.g.,][]{Clark+11,Greif+12,Dutta16,Raghuvanshi+23}. However, in reality, the fragments are surrounded by a massive dense region and quickly accrete matter from the surrounding and their mass increases within no time. Although this formation and instant growth of mass have been modelled by the sink particle, this is not feasible in the semi-analytical model. Therefore, we start our simulation from an epoch where the protostars have already been formed and have accreted a significant amount of mass. That is why we choose the initial mass ranging from $0.3 \ M_{\odot}$ to $0.7 \ M_{\odot}$.

Position coordinates in the radial direction within the $x-y$ plane and in the $z$ direction follow Gaussian distributions of $\mathcal{N}(100,10)$ and $\mathcal{N}(0,1)$, respectively. The distribution is uniform in the angular ($\phi$) direction. Radial (in $x-y$ plane), angular ($\phi$), and $z$ velocities (in $\mathrm{km/s}$) are sampled from $\mathcal{N}(5.5,3)$, $\mathcal{N}(5,3)$, and $\mathcal{N}(0,0.1)$, respectively. To incorporate both infalling and outgoing protostars, we randomly assign a positive or negative sign to the radial velocity after sampling from the distribution. These initial conditions (IC) closely align with previous studies \citep[see e.g.,][]{Machida+13,Dutta16}. Here we explore the time evolution only from a single IC. The statistical trends arising from different ICs are discussed in detail in Sec. \S\ref{sec:stat}.

The free parameters of the model we choose illustrate the physical characteristics of the system. For example, the impact of rotation of the gas has been denoted with $g_{\rm rot}\approx0.2$, which describes the ratio of the rotational velocity of the gas to the Keplerian velocity. The other parameters, i.e., $n_0\approx10^{13}\rm \ cm^{-3},r_0\approx5$ AU replicate the central dense regime of the cloud, which depends on the initial configuration of the system. The parameter $f\approx1$ defines the fraction of the incoming energy, which is being radiated by the protostar.

\section{Results} 
\label{sec:results}

To this effect, we evolve the system of protostar up to the epoch $t\approx10^6$ yr, which helps us investigate the long-term evolution of the system, including the dynamics of binary formation, merger phenomena etc from our IC. A detailed description of the same in case of different ICs has been illustrated in Sec. \S\ref{sec:stat}, showing that our results are consistent across various scenarios. Hereafter, the initial set of ten protostars will be designated as p1, p2, ..., and p10, respectively.

\subsection{Dynamics of the protostars} 
\label{sec:dynamics}

\begin{figure*}
\centering
\subfigure{\includegraphics [width=1.0\textwidth]{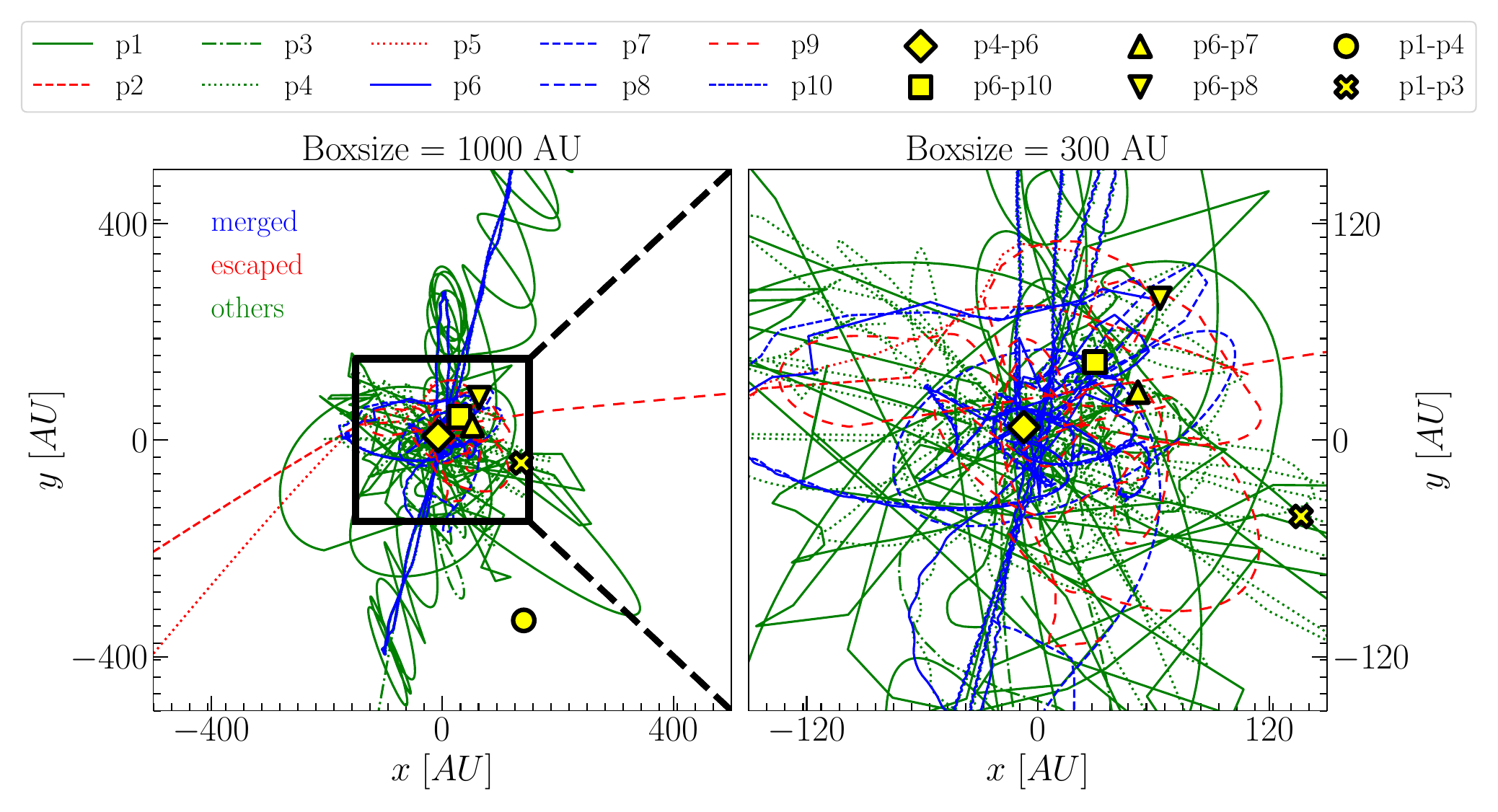}}
\caption{The trajectories of all the protostars in the $x-y$ plane have been visualised in the central regime of $\approx1000$ AU at an epoch $t\approx20.22$ kyr (left panel). The zoomed-in version (i.e., boxsize $\approx 300$ AU) on the right panel clearly depicts the dynamics of all the protostars that escape (denoted in red), merge (denoted in blue) and roam around (denoted in green), respectively. The yellow markers represent the four different sites where the merger event took place for the following protostars: the 1st merger took place at an epoch $t\approx0.0098$ kyr between p6 and p8 protostars, denoted by (p6-p8 pair). The 2nd, 3rd, 4th, 5th and the 6th took place at $t\approx8.81$ kyr for (p6-p10 pair), $t\approx9.6$ kyr for (p4-p6 pair), $t\approx19.68$ kyr for (p6-p7 pair), $t\approx146.7$ kyr for (p1-p4 pair) and $t\approx169.3$ kyr for (p1-p3 pair), respectively.}
\label{fig:xy}
\end{figure*}

Fig. \ref{fig:xy} shows the trajectories of all the protostars within the $x-y$ plane where the green color represents the protostars roaming around the cloud, blue color denotes the merged ones (described below) and the red color depicts the protostars moving away from the system, respectively. The left panel shows the overall dynamics of these protostars in the central regime of $\approx1000$ AU at an epoch $t\approx20.22$ kyr. The zoomed-in version of the same (shown in the right panel) is illustrated within a boxsize of $\approx300$ AU in order to visualise the trajectory in a closer examination.

Our simulation also captures the merging phenomenon between a few protostars that took place at six different epochs and six different sites, shown by the yellow markers. The first merging event happens between the protostars p6 and p8 (denoted by the ``p6-p8'') at an epoch $t\approx0.0098$ kyr, followed by the second between ``p6-p10 pair'' at $t\approx8.81$ kyr, the third between ``p4-p6'' at $t\approx9.6$ kyr, the fourth between ``p6-p7 pair'' at $t\approx19.68$ kyr, the fifth between ``p1-p4 pair'' at $t\approx146.78$ kyr and the sixth between ``p1-p3 pair'' at $t\approx169.3$ kyr, respectively. Interestingly, we also find that five of the six mergers take place within the central regime of $\approx150$ AU. This is due to the fact that the dynamical friction tends to be highest in the central dense part.

\subsection{Accretion phenomenon} 
\label{sub:dynamics}

\begin{figure}
    \centering
    \includegraphics[width=0.5\textwidth]{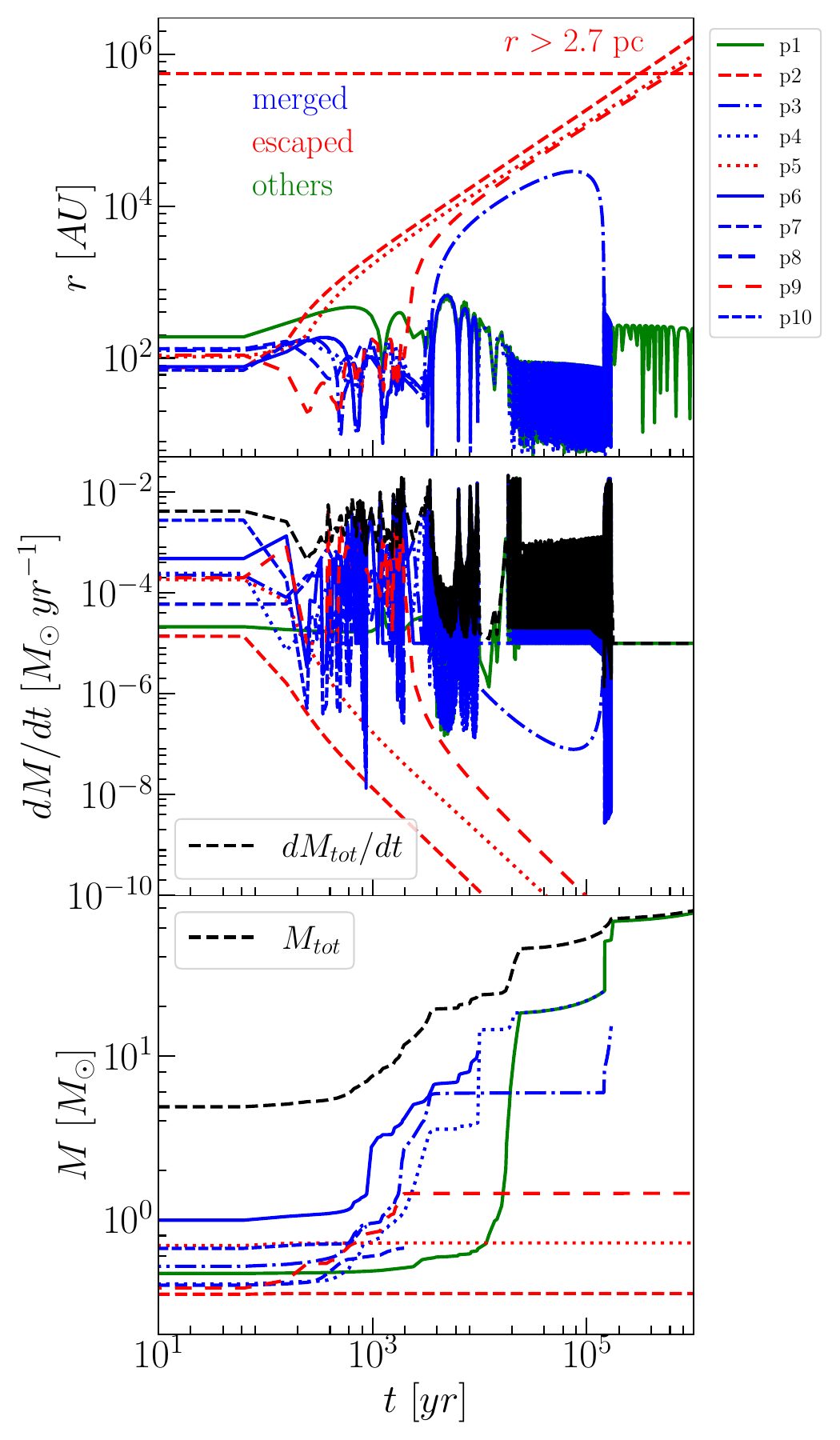}
    \caption{Mass accretion phenomenon has been demonstrated throughout the evolution epoch. The top panel denotes the possible candidates for the evolving protostars that may escape the cloud (denoted in red), merge with others (denoted in blue) and roam around within the cloud (denoted in green), respectively. Middle panel represents that the maximum accretion rate approaches a value $\approx 10^{-2}-10^{-3} \ M_{\odot} yr^{-1}$ for the protostars staying inside. This is  consistent with existing results from simulations \citep[see e.g.,][]{Abel:2001pr,Greif_2011,Dutta16,Latif_2022}. The black dashed line stands for the total accretion rate by all the protostars. The bottom panel shows the variation in the mass evolution in which we observe both low ($\lesssim2 \ M_{\odot}$) and high mass (in the range of $\approx6-75\ M_{\odot}$) protostars in our simulation. The high-mass protostars generally correspond to the ones that stay within and keep on accreting mass. The notation $M_{\mathrm{tot}}$, represented by the black dashed line, manifests the total mass in protostars.}
\label{fig:time_evolution}
\end{figure}

Here, we investigate the dynamical evolution of the accreting protostars, shown in Fig. \ref{fig:time_evolution}. The radial distance of the protostars shown in the top panel demonstrates three key features. First, we find that two candidates, namely protostars p2 and p5 are likely to move to the outer periphery straight away (red) without major interaction with the other protostars. Another protostar p9 (red) spends a significant amount of time ($t\approx2\times10^3$ yr) in the central regime before it gets ejected from the potential well of the gas following a close encounter with its counterparts. 
On the other hand, the blue ones have the possibility of having a close encounter, leading to a merger with other protostars. The protostar p3 particularly goes a large distance before it comes back again to the centre and keeps rotating, until it merges with protostar p1. Finally, among the protostars that are not likely to escape, we find only one protostar to survive (p1) at the end of the simulation (green), while all others end up being merged into it. There is also a possible formation of the binary system, described in section \S\ref{sub:binary}.

The mass accretion phenomenon has been demonstrated in the middle panel. The accretion rate for the protostars p2 and p5 continues to decrease monotonously as they move towards the lower density regime (i.e., the outer periphery) of the cloud. The protostar p9, on the other hand, shows a different trend of the mass accretion rate. This is because it spends a considerable amount of time in the central regime up to the epoch $t\approx10^3$ yr and then moves away to the outer periphery. All of these ejected protostars have an accretion rate of the order of $\approx10^{-8}-10^{-10}\ M_{\odot}yr^{-1}$ at the end of our simulation. On the other hand, the accretion rate for the non-escaping protostars (green, blue) shows strong fluctuations ($10^{-6}-10^{-2}\ M_{\odot}yr^{-1}$). This is because the close encounters assert rapid fluctuations in the velocities of these protostars, which get imprinted in the accretion rate. We denote the total accretion rate ($dM_{\mathrm{tot}}/dt$) by the black dashed line. 

The bottom panel illustrates the mass evolution of these protostars. Both p2 and p5, which are likely to be ejected, contain masses of $0.35\ M_{\odot}\ \mathrm{and}\ 0.72\ M_{\odot}$, respectively due to the low accretion rate. On the other hand, the mass of the protostar p9 ($M_{p9}\approx\ 1.44M_{\odot}$) that increases gradually at the initial stages becomes constant at $t\approx10^3$ yr as a result of the ejection. Among the others, we find p1 and p4 to evolve with similar masses $\approx24.2\ M_{\odot}$ upto $t\approx1.46\times10^5$ yr, followed by a merger. The protostar p3 too grows to $\approx15\ M_{\odot}$ before merging with p1.

\subsection{Time Evolution of mass function}
\label{sub:imf}

\begin{figure}
\centering
\subfigure{\includegraphics [width=0.2857\textwidth]{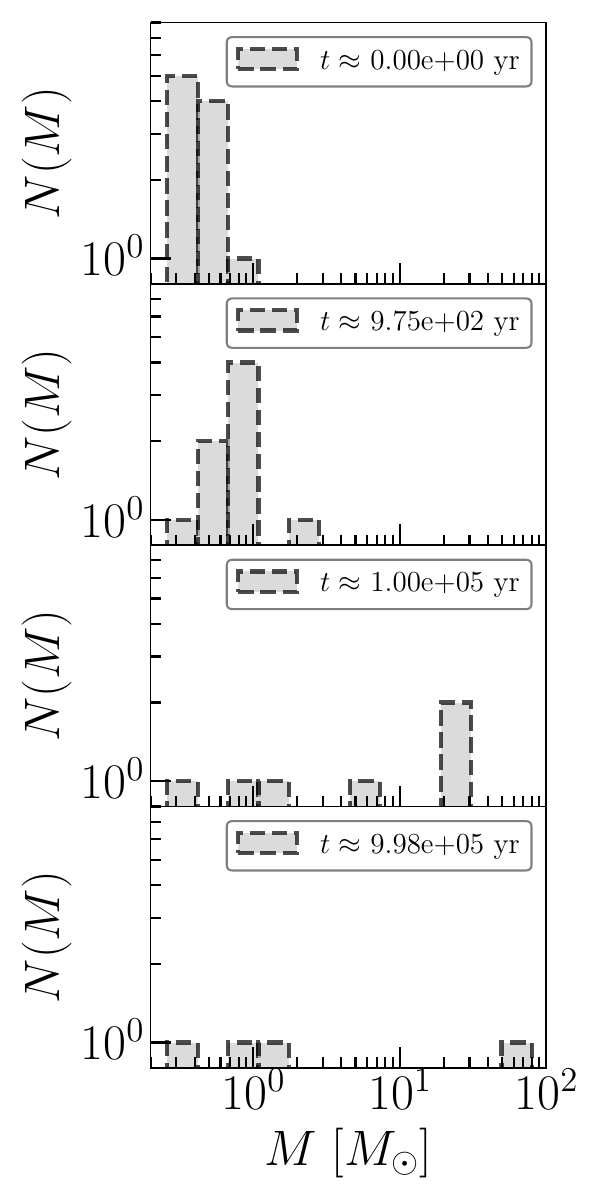}}
\caption{The time evolution of the mass function can be envisaged from four different epochs ($t\approx0,9.75\times10^2,1.0\times10^5,9.98\times10^5$ yr, respectively) from the top to the bottom panel. It is to be noted that the peaks of the mass function gradually migrate towards the higher end of the mass spectrum over time as a consequence of continuous mass accretion. This is expected as there is no formation of new protostars in our model.}
\label{fig:imf}
\end{figure}

Following the previous discussions, it is important to understand the distribution of the masses of the protostars as a function of time. This implies demonstrating the number of protostars in a particular mass range with the help of a histogram. Fig. \ref{fig:imf} illustrates how the mass function evolves with time. The four panels from the top to the bottom represent four different epochs $t\approx0,9.75\times10^2,1.00\times10^5,9.98\times10^5$ yr.

Initially, the mass function shows peaks only in the low-mass end, reflecting the initial configuration. Over time, the peaks of the mass function tend to shift towards the higher mass end as a result of the continuous accretion and merger by the protostars. This is expected since our model does not account for the formation of new protostars. Hence, at the end of our simulation, we find only one massive ($\gtrsim20 \ M_{\odot}$) and three low-to-intermediate-mass ($\lesssim 2 \ M_{\odot}$) protostars.

\subsection{Binary Formation}
\label{sub:binary}

\begin{figure}[htbp]
\centering
\subfigure{\includegraphics[width=0.45\textwidth]{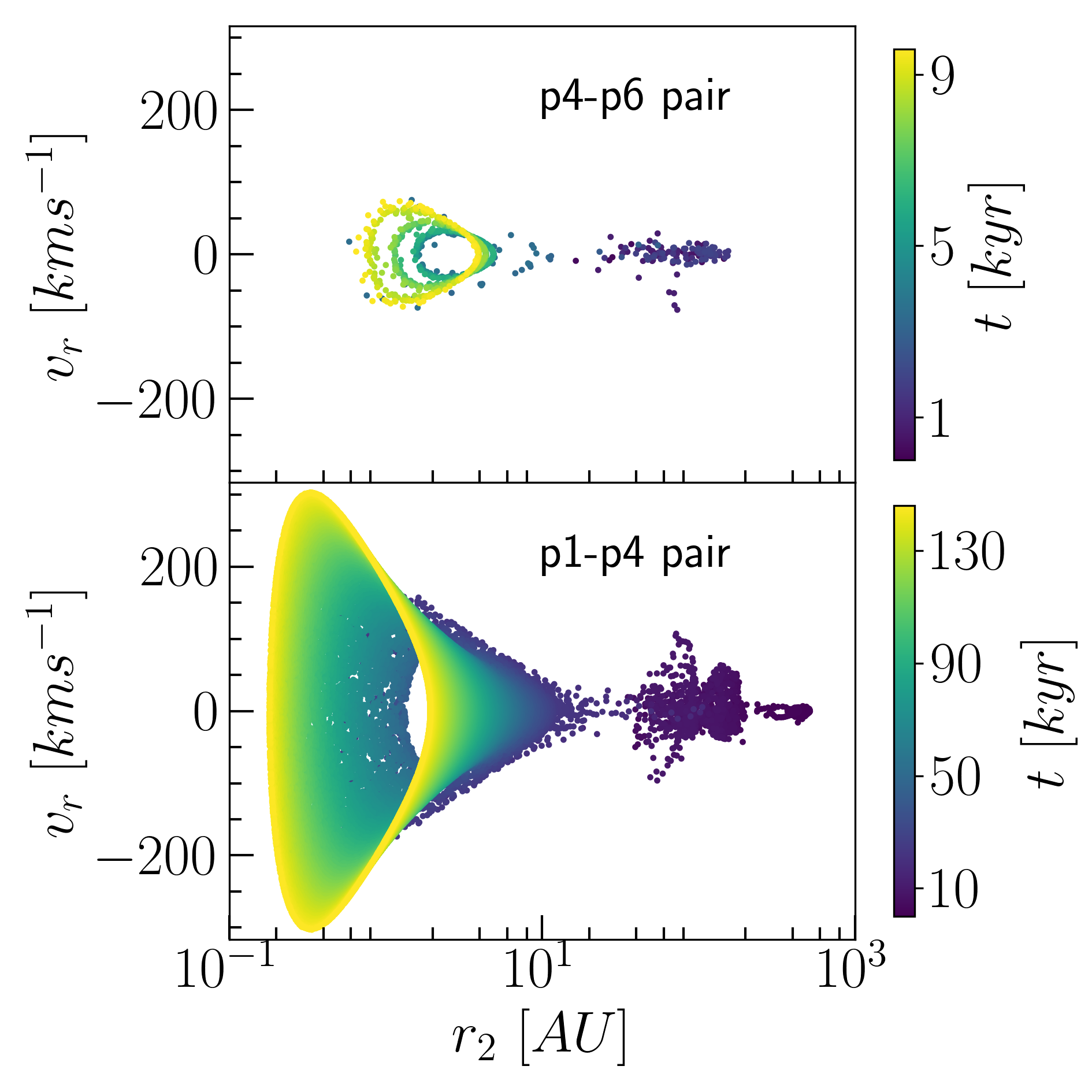}}
\caption{The phase space diagram ($r_2-v_{r}$ plot, where $r_2$ represents the distance between the pair) demonstrates the development of two binary systems in our simulation. The upper and lower panels markedly illustrate the formation of the first binary pair, denoted by ``p4-p6'', and the second binary pair, denoted by ``p1-p4'', respectively. The protostar p4 is likely to be merged with its counterpart p6 at $t\approx9.6$ kyr. Later, it forms another binary pair with the protostar p1 at an epoch close to 40 kyr. The second pair also seems to end up merged at $t\approx146$ kyr.}
\label{fig:binary}
\end{figure}

The implementation of the N-Body dynamics allows us to capture the formation of binary systems in our simulation. We demonstrate the formation of two different binary pairs in Fig. \ref{fig:binary}. 

The phase space diagrams, i.e., the $r_2-v_r$ plot, where $r_2$ and $v_r$ signify the distance and relative radial velocity between the protostars, imply the development of an ordered and bounded state over time. The top panel shows that the first binary pair forms between the protostars p4 and p6, denoted by the ``p4-p6 pair'' at $r_2\approx10$ AU around an epoch $t\approx5$ kyr. We observe that the evolution of this ``p4-p6 pair'' appears to be terminated at an epoch $t\approx9.6$ kyr. This is because the protostar p6 is likely to be merged with protostar p4, which is going to form another binary system with the protostar p1, at a later epoch of time (bottom panel). The second one, i.e., the ``p1-p4 pair'', which tends to build at a distance $r_2\approx25$ AU around the epoch $t\approx 40$ kyr, continues to evolve till $t\approx146$ kyr when it seems to get merged into a single protostar.. 

The imprints of the binary formation can also be visualised on the dynamics of the protostars shown in the top panel of Fig. \ref{fig:time_evolution} in the previous section \S\ref{sub:dynamics}. There we see that the protostar p6 has a similar trajectory with protostar p4 during the period $\approx10^3-10^4$ yr, indicating the formation of the first binary system. In the same plot, we see that as time progresses, the trajectory of the protostar p4 becomes close to that of the protostar p1 during the epoch $\approx2\times10^4\ \mathrm{yr}-10^5$ yr. This indicates the development of the second pair between p1 and p4. Additionally, we find wide oscillations in the accretion rate ($10^{-2}-10^{-7} M_{\odot}yr^{-1}$) for these protostars (shown in the middle panel of Fig. \ref{fig:time_evolution}). This is a result of rapid changes in their velocities that are caused by strong interactions.

\subsection{Possible ejection of the protostars}
\label{sub:fd}

\begin{figure}
\centering
\subfigure{\includegraphics [width=0.43\textwidth]{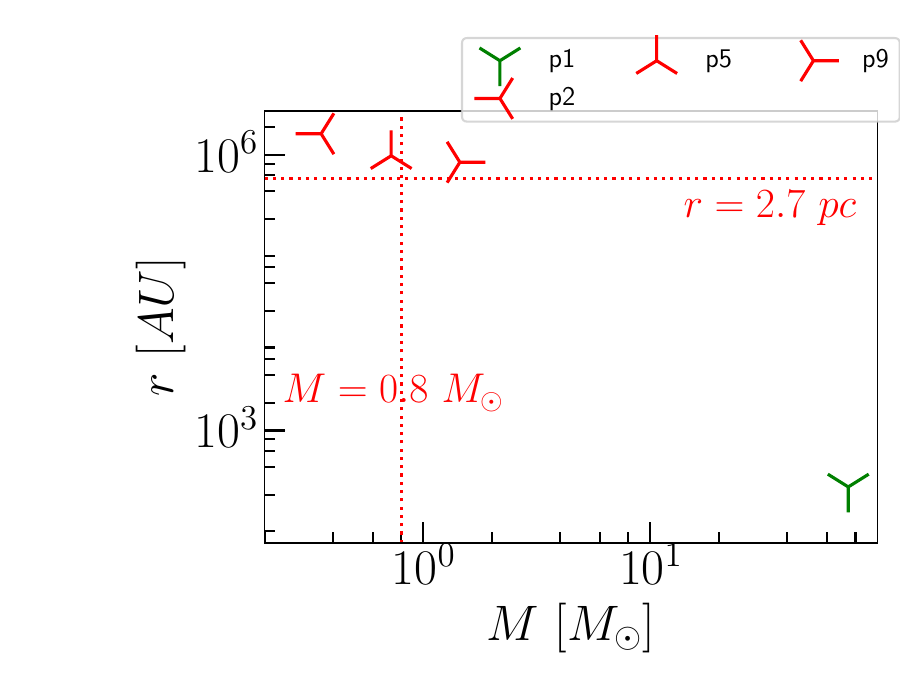}}
\caption{Mass and radial distance of the four protostars that are left over (out of the ten at the beginning of our simulation) as a consequence of the merger events, are plotted at the end of our simulation. The scatter diagram shows that three protostars are likely to move away (denoted by red marker) from the potential well of the cloud. It is to be noted that the ejected protostars accommodate two low-mass protostars ($M_{p2}\approx0.35\ M_{\odot}$ and $M_{p5}\approx0.72\ M_{\odot}$) and another protostar with a slightly higher mass ($M_{p9}\approx1.44\ M_{\odot}$), respectively. These low-mass protostars may possibly go through the ZAMS phase and hence remain as the probable candidates to survive for a longer period of time. The other one (denoted by green markers) seems to remain inside the cloud as a high-mass protostar.}
\label{fig:sum}
\end{figure}

Following the merging phenomenon, our final snapshot at $t\approx10^6$ yr contains nearly four protostars of different masses that are distributed throughout the cloud. The evolution of the mass function of these protostars has already been discussed in the previous section. Here we describe the possible implications of our findings.  

The scatter diagram in Fig. \ref{fig:sum} denotes that three protostars contain an effective radial velocity that exceeds the escape velocity of the system. Hence, there is a possibility for these protostars to get ejected from the cloud. We observe that two of the ejected protostars contain low mass ($M_{p2}\approx0.35\ M_{\odot}$ and $M_{p5}\approx0.72\ M_{\odot}$), while another one tends to have a slightly higher mass ($M_{p9}\approx\ 1.44\ M_{\odot}$). According to the theoretical perception of stellar evolution, massive stars have a shorter lifespan and stars with a mass $\lesssim0.8\ M_{\odot}$ (shown in the red dashed line) have the possibility to enter the zero-age-main-sequence (ZAMS) phase. In view of this consensus, we can presume that both the low-mass protostars, i.e., p2 and p5, are likely to enter the ZAMS phase, and hence become the probable candidates to be survived for a longer period of time. 

The other protostar appears to contain more mass ($M_{p1}\approx74.24\ M_{\odot}$) during the course of evolution. These high-mass stars may either develop the seed of the black holes or explode as PISN supernovas depending on the accretion phenomenon and other physical properties of the system.

\section{Effects of the free parameters}
\label{sec:stability}
Till now, we have shown our results for a chosen set of parameters (i.e., $f,g_{\mathrm{rot}},n_0,r_0$), which determine the physical properties at some particular epoch. However, in reality, all these parameters evolve with time, which is a challenging task due to its complexity to simulate in a semi-analytical model. In this section, we try to overcome this difficulty by exploring the possible effects that may be induced by the variation in these free parameters as a first approximation.

\subsection{Variation of the strength of feedback}
\label{sub:imf_f}

\begin{figure}
\centering
\subfigure{\includegraphics [width=0.29\textwidth]{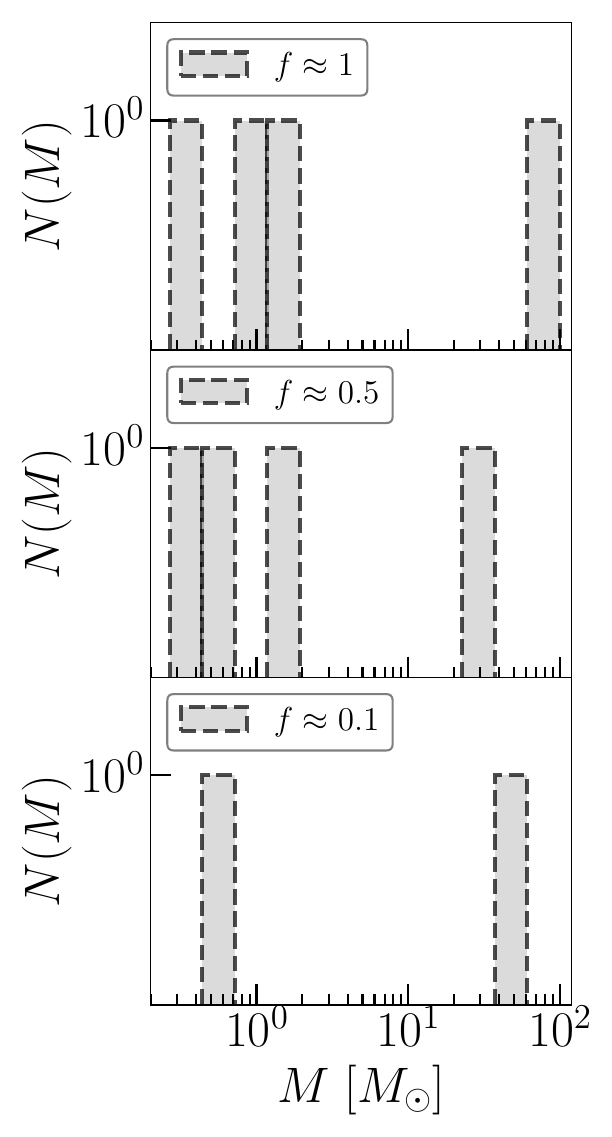}}
\caption{The mass function is depicted for three distinct values of the parameter $f\approx1,0.5,0.1$, which represents the fraction of the incoming energy radiated away by the protostar, at the end of the simulation. The lower value (i.e., $f\approx0.1$ in bottom panel) corresponds to the existence of only two protostars. This is because the dynamical friction leads to more inward dragging of the protostars, consequently resulting in an increased number of mergers. In contrast, as $f$ approaches higher values ($f\approx0.5,1$, in the middle and top panel, respectively), we observe a comparatively higher number of protostars in our simulation.}
\label{fig:imf_f}
\end{figure}

As discussed before, the parameter $f$ represents the fraction of the incoming energy radiated away by the protostars, heralding the strength of the feedback. Hence, a lower value of $f$ ensures more accretion onto the protostar for a longer period of time. This results in an increase in the dynamical friction, which drags the protostars towards the central regime. As a consequence, the possibility that the protostars merge with each other seems to increase. Thereby, the less is the strength of feedback, the more is supposed to be the rate of merger.

Fig. \ref{fig:imf_f} conclusively demonstrates this effect, where the top, the middle and the bottom panel correspond to the mass function at the final snapshot for $f\approx1,0.5,0.1$, respectively. Both the top and the middle panels display four protostars, while the bottom panel shows only two, indicating an increased likelihood of mergers. The existence of only one low-mass protostar compared to higher values of $f$ also demonstrates the fact that the escape fraction is likely to be reduced by the dynamical friction.

\subsection{Variation of cloud properties and the suppressed accretion rate}
\label{sub:vary}

Throughout our work, we have assumed a static background, i.e., the parameters $n_0,r_0,g_{\mathrm{rot}}$, which represent the configuration of the cloud, remain unchanged. However, in reality, the size, density, temperature etc of the core varies continuously with time \citep{Yoshida_08}. Because our model does not have this effect built-in, we vary the above parameters one by one in order to have an estimate of the impact on the final results. Moreover, as the cloud undergoes collapse, the rotation of the gas also evolves with time.
	
\begin{figure}
\centering
\subfigure{\includegraphics [width=0.4\textwidth]{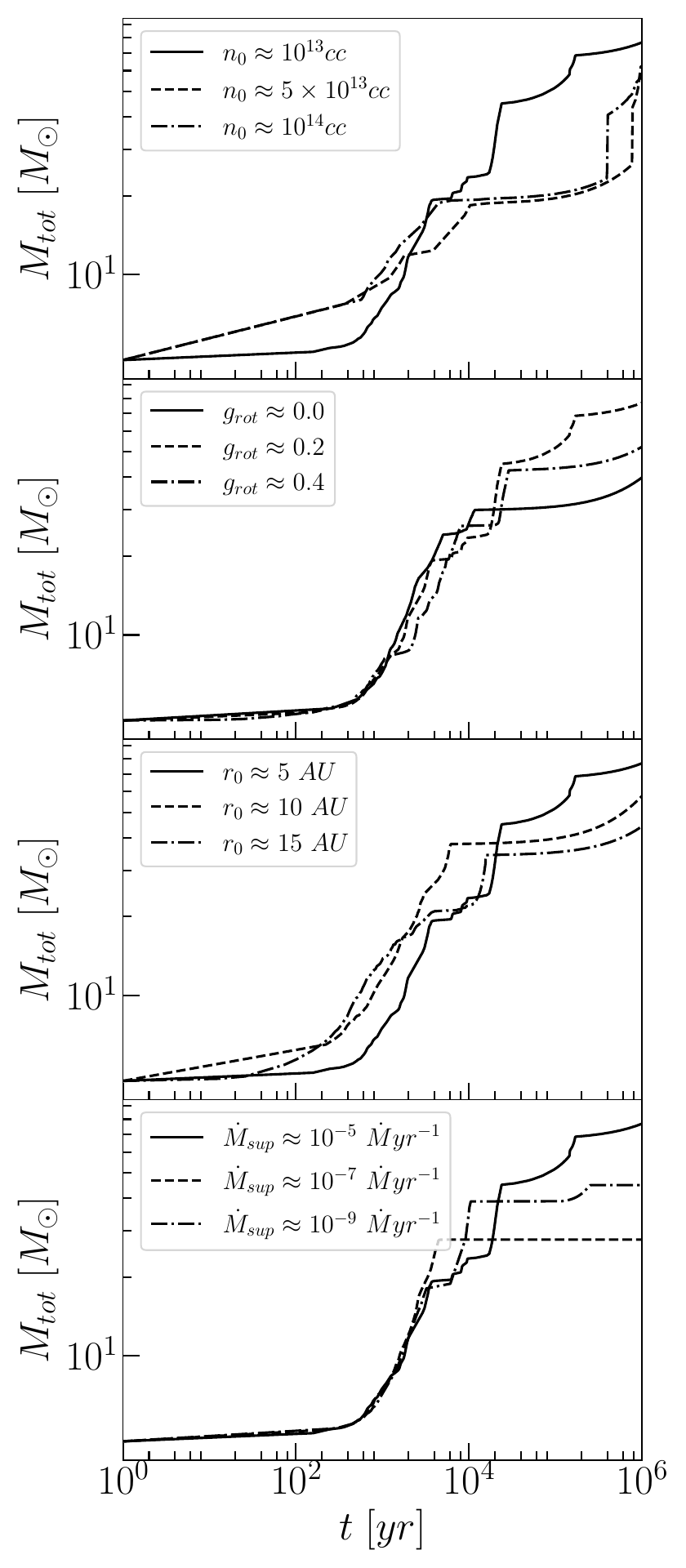}}
\caption{The top panel illustrates the fact that the total mass in all the protostars is likely to increase at a comparatively higher rate for the denser cloud in the initial stage of evolution. However, this trend reverses at around $t\approx3-4$ kyr, once the radiation pressure becomes significant and starts impacting the mass accretion rate. We observe a similar characteristic in the second panel that shows the effects of gas rotation ($v_{\mathrm{gas}}$) on the mass evolution. This is represented by the three rotation parameters $g_{\mathrm{rot}}\approx0.0,0.2,0.4$. As the mass accretion rate is inversely proportional to the relative velocity between the gas and the protostars, a fast-rotating cloud results in the formation of comparatively low-mass stars. Once again, feedback changes the scenario in the later stages. The third panel shows the impact of the core size on the total mass evolution, which follows a similar trend until the epoch when radiative feedback becomes significant. Finally, the bottom panel demonstrates the effects of varying the suppressed accretion rate $\dot{M}_{\mathrm{sup}}$ on the time evolution of the total mass in the protostars. We do not find the total mass to change dramatically as we vary the suppressed accretion rate.}
\label{fig:vary}
\end{figure}

The top panel shows that the total mass of all the protostars within a comparatively denser cloud increases rapidly at the initial stages of evolution. Interestingly, the trend reverses at a later stage $t\approx3-4$ kyr. This is due to the fact that the radiation pressure starts impacting the mass accretion process at this epoch \citep[shown in earlier studies, see e.g.,][]{Stacy+12,Latif_2022}. This happens because the Eddington limit is reached at an earlier time when the core is more dense. This also implies the formation of higher mass protostars in a cloud corresponding to lower central density.

From the second panel, we notice that the mass accumulation at the initial stages is generally higher in the case of slow-rotating clouds that is represented by the rotation parameter $g_{\mathrm{rot}}\approx0.0,0.2$. This happens because the accretion rate is inversely proportional to the relative velocity ($v_{\mathrm{rel}}$) of the protostar with respect to the gas, i.e., $\dot{M}\propto (1/v_{rel}^3)$, where  $v_{\mathrm{rel}}\approx\vec{v}-\vec{v}_{\mathrm{gas}}$. Therefore, a high rotating cloud ($g_{\mathrm{rot}}\approx0.4$) results in the formation of low-mass protostars. When the feedback comes into action at a later stage around $t\approx3-4$ kyr, the trend reverses again as in the previous case. 

The size of the core could also have a significant impact on the mass of the protostars. For example, protostars within a cloud of a larger core (such as $r_0\approx15$ AU) accumulate more mass up to the epoch $t\approx3-4$ kyr, as shown in the third panel. Similar to the earlier trends, the behaviour seems to become the reverse due to the influence of the feedback process.

Finally, the bottom panel describes how the total mass is affected by variation in the suppressed accretion rate $\dot{M}_{\mathrm{sup}}$. The total mass seems to be higher by a factor $\approx 3$ at the largest value of $\dot{M}_{\rm sup}\approx 10^{-5}M_{\odot} yr^{-1}$. However, for $\dot{M}_{\rm sup}\approx 10^{-7}\ M_{\odot} yr^{-1}$ and $\dot{M}_{\rm sup} \ \approx 10^{-9}\ M_{\odot} yr^{-1}$, more mass is accumulated in the latter case. The initial evolution does not change; but this is expected since the differences would only arise once the Bondi-Hoyle accretion rate exceeds the Eddington condition. Together, these four diagrams may justify the stability of our numerical model.

\section{Statistical trends from five different random realizations (ICs)}\label{sec:stat}

Till now, our calculation is based on a single initial condition (IC) and is just dependent on changing the model parameters. However, the key advantage of the presented scheme, i.e., the cost, is its potential for exploring several realizations of initial conditions. This is necessary to scrutinize whether the trends or features are susceptible to stochastic effects or not. Simulating with one IC makes it unclear to what extent they are true trends or just statistical noise imposed by the IC. Hence, we carry out an identical set of calculations for four other different random realizations of the ICs following the same prescription to pick out the more general trends in the results.

\subsection{Different initial conditions}

We generate four random realizations, denoted by `r1', `r2', `r3', and `r4' along with the fiducial model. The distribution of the protostars' mass, radial distance, and the three components of velocity, namely `$v_r$', `$v_\phi$', and `$v_z$' is illustrated in Fig. \ref{fig:ic_all}. The values of the parameters in all five cases are in accordance with the previous studies. The details have been elaborated in section \S\ref{sub:ic}. In all five cases, the system is evolved for nearly a million years ($t\approx10^6$ yr).

\subsection{Comparing the evolution of radial distance, accretion rate and mass of protostars for five different ICs}

The time evolution of the radial distance, accretion rate and mass of these protostars in each realization is depicted in Fig. \ref{fig:all}. The free parameters ($f\approx1$, $r_0\approx5$ AU and $n_0\approx10^{13}\rm \ cm^{-3}$) are similar to our fiducial model. 
 
 (i) We find that three protostars are likely to escape in the case of the fiducial run and realization `r1'. In realization `r2', the number of escaping protostars is four whereas for realization `r3' and `r4', it is two. 
 
 (ii) In all cases, we observe only one massive protostar near the central regime of the cloud. 
 
 (iii) The existence of protostars marked with blue suggests the presence of the merger phenomenon in all realizations.
 
 (iv) The mass accretion rate of the escaping protostars drops in all cases, where the non-escaping protostars show an accretion rate up to $\approx10^{-2}\ M_{\odot}yr^{-1}$ in each case. 
 
  (v) The total mass in the protostars for all different realizations shows a trend of a consistent range of $\approx 30-76\ M_{\odot}$. The above features also point towards the stability of our semi-analytical model.

\subsection{Comparison of Binary Formation in all five random realizations}

 Interestingly, we find the formation of binary pairs for all different random realizations, as depicted in Fig. \ref{fig:binary_all}, where we have demonstrated two binary pairs from each realization. The fiducial model shows two pairs between protostar 4 and protostar 6 (denoted as the `p4-p6' pair) and the `p1-p4' pair. Similarly, the realization `r1' contains the `p3-p8' and `p2-p3' pair, 'r2' contains the `p1-p6' and `p1-p2' pair, `r3' contains the `p1-p5' and `p1-p2' pair and finally `r4' contains the `p3-p5' and `p1-p2' pair. Noticeably, at the end of each simulation, all these binary pairs are likely to merge with each other. We emphasize that our protostars are accreting either at a Bondi-Hoyle rate or at a suppressed accretion rate $\dot{M}_{\mathrm{sup}}$ during the entire time of evolution. Therefore they are subjected to a drag induced by the dynamical friction throughout their trajectories. Hence, it is quite expected that they are likely to merge with each other at some epoch of time. This trend holds for all five random realizations.

\subsection{Finding statistical features with the radiation factor f}

Now we explore how the change in the radiation factor $f$ impacts the different aspects of the dynamics of the protostars, such as mergers, IMF etc. This is particularly important because this factor controls the onset of the maximum accretion rate induced by the radiation pressure, which determines both the accretion and dynamical friction. This can possibly influence the physical processes like the merger and the escape phenomenon. In addition, the maximum mass of the protostars can also be constrained by the factor $f$.

\subsubsection{Merger}

\begin{figure}[htp]
    \centering
    \subfigure[]{
        \includegraphics[width=0.48\textwidth]{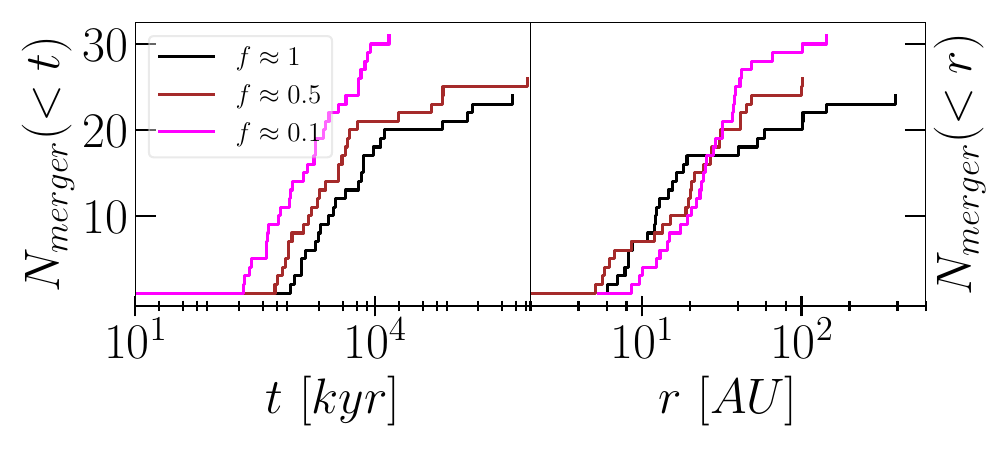}
    }
    \caption{The left panel demonstrates the cumulative number of mergers with time taking all five random realizations into account for three values of $f\approx0.1,0.5,1$. As expected, a lower value of $f$ leads to a higher number of mergers that kick in earlier as a result of dynamical friction. On the other hand, we do not see such a distinct trend in the cumulative number of mergers with the radial distance in the right panel.}
    \label{fig:merge}
\end{figure}

Fig. \ref{fig:merge} demonstrates the imprints of varying the radiation factor $f$ on the merger phenomenon. The left panel shows the cumulative number of mergers as a function of time in all five random realizations for three values of the radiation factor $f\approx1,0.5,0.1$. It is to be noted that a lower value of $f$ results in more mergers, which also seems to commence earlier. We know that the highest accretion rate increases with a lower value of $f$, as a result of radiation pressure. Hence, we attribute this increased rate of merger to the higher amount of drag on the protostar. On the other hand, we observe no such prominent trend in the cumulative radial distribution from the right panel of Fig. \ref{fig:merge}.

\subsubsection{Mass - radial distance diagram}

\begin{figure*}[htp]
    \centering
    \subfigure[b][The protostars end up either in the outskirts of the cloud with a low mass, or near the centre as massive.]{
        \includegraphics[width=0.4\textwidth]{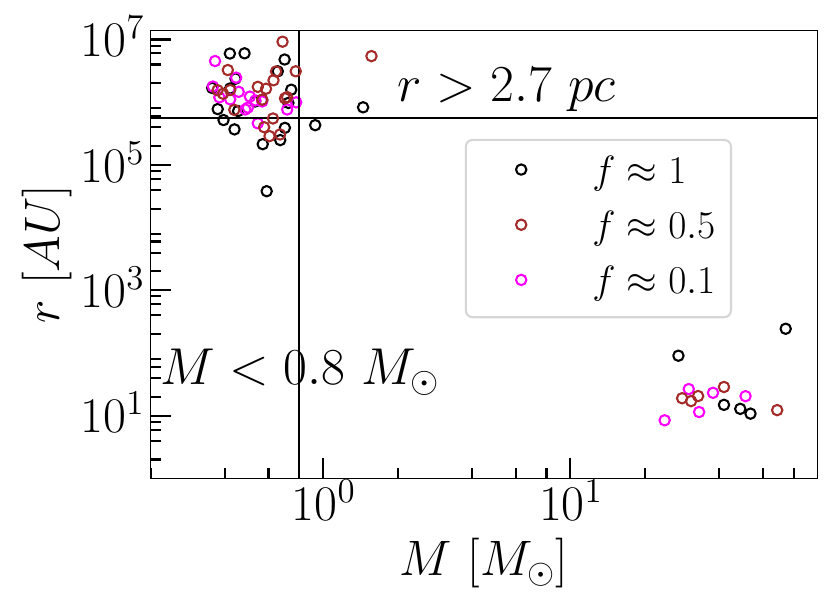}
        \label{fig:mr_f}
    }
    \hspace{1cm}
    \subfigure[b][Distribution of both the mass (left panel) and radial distance (right panel) of the protostars show dual-peak nature.]{
        \includegraphics[width=0.4\textwidth]{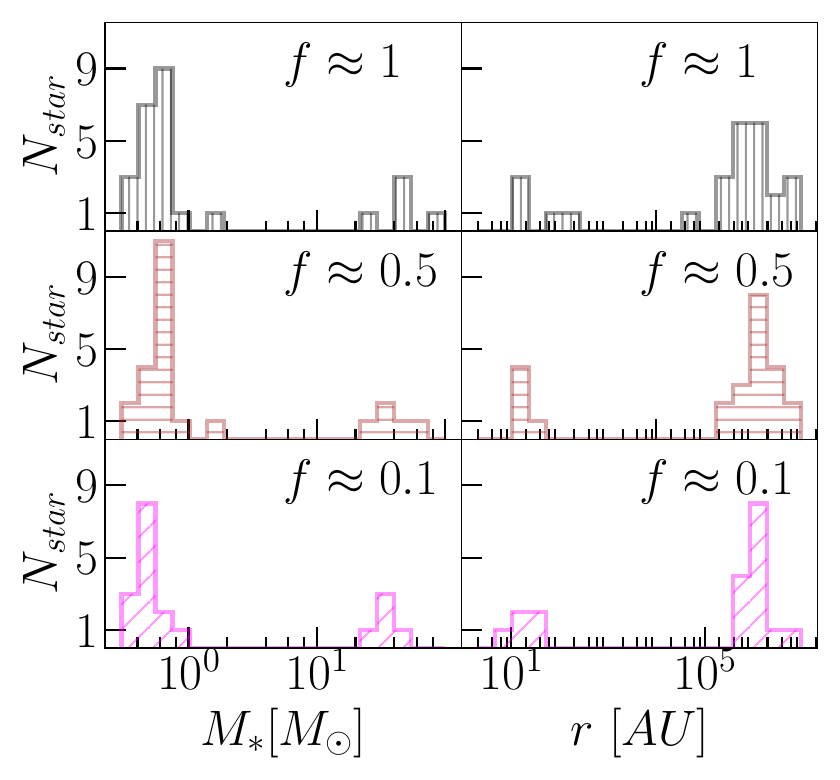}
        \label{fig:imr}
    }
\caption{Combined scatter distribution of the mass and radial distance of the protostars from all five random realizations is demonstrated in the left panel, whereas the right panel depicts their individual distributions. Both the trends hold for all three values of $f\approx1,0.5,0.1$.}
\end{figure*}

Fig. \ref{fig:mr_f} shows the final distribution of the mass and radial distribution of the protostars for three different values of $f\approx0.1,0.5,1$. In each case, we find the protostars of low-mass $\lesssim1\ M_{\odot}$) at the outskirts of the cloud or high-mass ($\gtrsim20\ M_{\odot}$) protostars near the central regime. The possible candidates for the protostars to be survived till the present epoch of time can be observed in the top left regime, limited by the radial distance $r \geq 2.7$ pc and mass $M \lesssim 0.8\ M_{\odot}$. We find five high-mass protostars for each value of $f$, corresponding to five different realizations. This re-instates our previous observation that only one protostar ends up as a massive one at the end of each simulation. The individual distribution of the mass and radial distance of the protostars is depicted in Fig. \ref{fig:imr} that reconfirms the dual-peak nature of their distributions.

Contrary to expectation, higher values of $f$ do not necessarily produce lower-mass protostars, as shown in Fig. \ref{fig:mr_f}. As we have seen earlier, higher values of $f$ produce fewer mergers and hence we have more number of protostars in the system. As a result, the protostars go through more dynamical N-Body interactions, which bring about rapid changes in velocity. This in turn brings the Bondi-Hoyle accretion rate at a value lower than the maximum accretion limit. This enables the protostar to go through sustained accretion in an episodic way.

\begin{figure}[htp]
    \centering
        \includegraphics[width=0.4\textwidth]{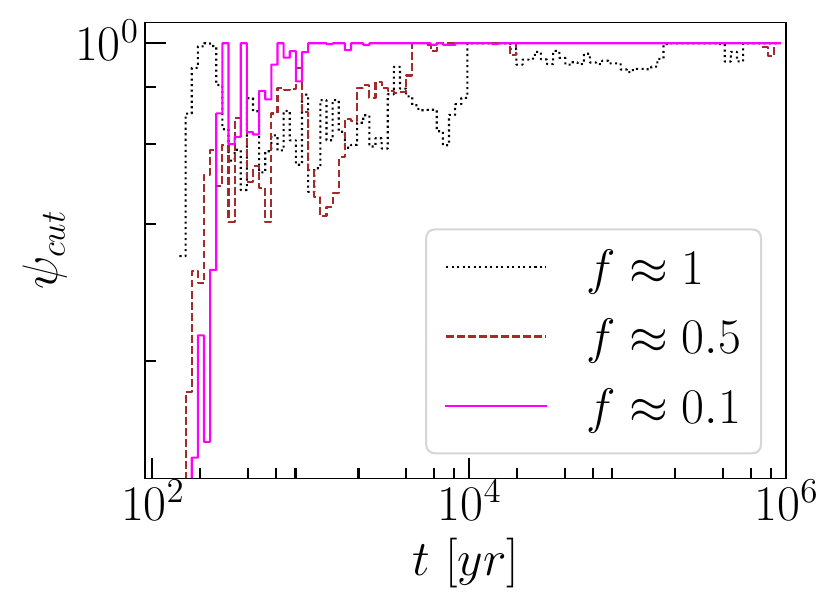}
    \caption{The likelihood of finding at least one protostar ($\psi_{\mathrm{cut}}$) whose Bondi-Hoyle accretion rate exceeds the maximum accretion rate cutoff is lower when the radiation factor $f$ is high. In such a scenario, fewer mergers lead to more protostars in the system. This introduces increased dynamical interactions that induce rapid fluctuations in their velocities. Thereby, the Bondi-Hoyle accretion rate  drops below the cutoff, resulting in a more sustained accretion.} 
    \label{fig:fbk}
\end{figure}

This is conclusively illustrated in Fig. \ref{fig:fbk}, which depicts how the probability of finding at least one protostar above the maximum accretion rate cutoff, varies with time. Clearly, it is less likely to find such a protostar when $f$ is high. The complete timespan of $t\approx10^6$ years is divided into 8000 bins, which have been spaced logarithmically. In each bin, we calculate (i) the total number of snapshots and (ii) the number of snapshots with at least one protostar above the maximum accretion rate cutoff. Finally, we use their ratio to calculate the factor $\psi_{\mathrm{cut}}$ in each bin.

Finally, after a rigorous investigation with different initial conditions, we are now in a position to conclude that all these physical phenomena emerging out of our model follow a consistent profile and are not merely a result of statistical noise. This implies that they are unsusceptible to the initial configuration of the system.


\section{Discussions} 
\label{sec:discussions}

As a continuation of an earlier work by \citet{Dutta+2020}, we have substantially improved the previous model in order to study the long-term evolution of the protostars. First we discuss the limitations and scope for improvements in \S\ref{sub:caveats} and the observational implications in \S\ref{sub:obs}, followed by a brief review on the applicability of our model in other astrophysical systems in \S\ref{sub:applicability}. Subsequently, we summarize the key points of our work in \S\ref{sub:key}.

\subsection{Limitations and scopes for improvement}
\label{sub:caveats}

(i) As we know, the collapse of a rotating spherically symmetric cloud of primordial gas results in the development of a rotating circumstellar disk. Due to the high complexity of the process that incorporates the simultaneous heating, cooling, radiation, chemical network and accretion; it is quite difficult to capture the entire evolution in a semi-analytical model. This is the reason we are motivated to develop a sophisticated model to see the long-term evolution. Notwithstanding, our attempt to model the effects of varying the core radius, density and gas rotation tries to simulate this intricate effect in a simplified setup, which may be considered as an alternative to traditional hydrodynamical simulations. Having said that, we are also in the process of including exponential disk in our model in the near future.

(ii) We assumed a static background in the current numerical setup in which the ambient density and the temperature remained unchanged throughout our simulation. In a 3D numerical simulation, the entire evolution process starting from the collapse to the formation of the protostars and the other properties such as the temperature, density, accretion and primordial chemistry can be captured, but for a few thousands of years. However, to investigate the fate of the protostar, the system needs to be evolved up to a million years, which is not feasible in a 3D hydrodynamical simulation till now. Hence, a semi-analytical model can address this issue with some approximations regarding the collapsed gas density. For example, most of the accretion occurs ($c_s^3/G$) in the central dense regime where the change of density and temperature varies the most, whereas the density and temperature in the outskirts remain nearly unchanged (See Fig. 2, \citep{Abel:2001pr}). In another assumption, the growing protostars in low-rotating clouds also condense in the central regime heralding the nearly static background in the outer periphery. With this approximation, one can go ahead with a semi-analytical model with a motivation to only focus on the evolution of protostars for a longer period of time. During the accretion process by the protostars, the gas in the local volume is depleted from the medium and hence loses energy. As a result, both the density and temperature change with time in that region. The same has been seen once the disk is formed. On the other hand, this variation in the background profile is negligible in the outer periphery and hence can be summed to remain in the steady state. To summarise, the protostars in the outer periphery are not likely to be directly affected by the central dense regime, whereas the others are evolved with few approximations.

(iii) The feedback-accretion scenario has been implemented assuming spherical symmetry in our model. However, this is almost certainly not the case in reality and the accretion follows a highly non-spherical process. Implementing an improved framework that can also mimic the non-spherical accretion processes will only aid our model.

(iv) We have used radiation pressure as the only channel in our model  \citep{Wolfire+87,Krumholz+09} that may not be representative of the most relevant channel for limiting the growth of massive Pop III massive stars.  Moreover, the feedback effects on the gas are also not captured, which is another indirect channel through which low-mass stars could be impacted by the feedback. This is more feasible in a 3D simulation. In our next work, we aim to include both radiation pressure and photoionisation effects \citep{Sales+14} to have a more realistic scenario.

In spite of these limitations, it is important to note that our feedback model captured the highest mass of the order of 24 $M_{\odot}$ preceding the merger, followed by a single merged protostar that evolves to $74 \ M_{\odot}$ at the end of the simulation. This result is similar to the finding \footnote{To be noted, the feedback mechanism that sets the upper limit in these simulations is photoionisation of the cloud, that halts accretion.} by \citet{Hosokawa+11} and \citet{Stacy+12} that says the photoionisation feedback impacts the protostellar accretion only when the protostar grows to a mass of $20-40\ M_{\odot}$. Hence, we do not expect the qualitative features of the mass function, e.g., the peak positions to change dramatically when both photoionisation and radiation pressure act together. Nevertheless, implementing a more realistic model for the feedback will only aid our results, which we leave for future work.

\subsection{Observational Implications}
\label{sub:obs}

Several state-of-the-art investigations have been carried out with a strong emphasis on the metal-poor stars within the galactic halo in order to detect the candidate for the Pop III stars \citep[see e.g.,][]{Salvadori+10,caffau+11,David+13,Frebel_Norris_15,Hartwig+15,Komiya+15,Else+17}. Besides, a series of recent investigations \citep[see e.g.,][]{Placco+15,Ishigaki_2018,Skuladottir_2021} on the chemical abundances of extremely-metal poor stars concluded that the Pop III stars could be of the order of $20-30\ M_{\odot}$, which aligns well with our results.

Moreover, in the present era, there has been a lot of attention on theoretical studies regarding the survival possibility of Pop III stars. That is why observations are more interested in finding the extremely-metal-poor (EMP) stars of low mass, that may be the possible candidates for the first generation of stars \citep[see e.g.,][]{Marigo+01,Komiya+16,Ishiyama_2016,Kirihara+19}. 

Notably, the Hubble telescope detected a very distant star \citep{Welch+22,Schauer+22} in the early universe. Recent JWST results \citep[see e.g.,][]{vanzella+23, Yajima+23} also indicate the possible detection of Pop III stars. A recent theoretical study by \citep{Venditti+23} suggests that the first generation of stars might have formed at a comparatively lower redshift of $z \approx 6-8$. This may be explored with the help of JWST deep surveys. All of these studies make it very interesting to motivate theoretical models to understand the evolution of the very first stars in the universe. This also highlights the significance of our work.

\subsection{Applicability in other astrophysical systems}
\label{sub:applicability}

The generalistic feature of our model is to investigate the long-term evolution of any multi (N)-particle system inside a supersonic and compressible flow of fluid, which evolves under gravity. In addition, our model also includes the physical processes such as accretion, merger phenomenon and radiative feedback. The initial condition (IC) of an N-particle system combined with the background profile such as density and temperature distribution of the fluid, in general, characterises a particular system.

For example, in this particular work, we have used the IC and the background profile for a collapsed primordial cloud where the mass accretion follows a Bondi-Hoyle description and the feedback is modelled using the radiation pressure. Below we mention some of the other cases where our model can be useful.

(i) The formation of supermassive black hole (SMBH) seeds is another topic of active research where our model can be used to study the runaway merger of multiple protostars that can possibly give rise to the seeds of SMBH \citep{Zwart+02,Vergara21}.

(ii) Currently our model is capable of handling only radiation pressure, as a tool of feedback mechanism, in the case of a spherical accretion scenario. Nevertheless, this model can be certainly extended to study the complex processes of radiative feedback such as photoionization, photodissociation etc and their effects on the accretion scheme \citep{Hosokawa+11,Jaura+22}.

(iii) Our framework can also be useful to study the protoplanetary systems where the protoplanetary disk fragments to form multiple protoplanets which grow through continuous accretion and mergers \citep{Ogihara+09,Michiel+19}.

  In conclusion, we would like to emphasize that this particular model can be used to study the lagrangian evolution of any multi-particle gravitating system where the time evolution of the physical properties can be estimated through a system of coupled ODEs.

\subsection{Key Points and conclusions}
\label{sub:key}

(i) To model a more realistic scenario, we introduced (a) rotation in the cloud, (b) N-Body interaction between multiple protostars, (c) radiation pressure as a form of the feedback process, (d) a model for the radius of the protostar instead of point mass particle that allowed us to study (e) the merging phenomenon.

(ii) The configuration of the protostars has been initialised with a uniform distribution of mass, whereas, their positions and velocities are drawn from the Gaussian distribution. 

(iii) We ensured that the angular velocity of the ambient medium closely follows the characteristics of a collapsed gas cloud. Besides, the escape velocity has been estimated analytically with a fitting function \citep[closely matches with][]{Dutta+2020}.

(iv) The trajectories of the N protostars have been distinctly visualised in a \textit{zoomed-in version}, where the merging phenomenon, accretion history and entire dynamics have been captured very precisely.

(v) The Mass accretion phenomenon has been considerably improved by including the well-known Eddington limit that helps us to understand mass evolution in detail.

(vi) One important aspect of our model is that it is capable of capturing the formation of binary pairs at different epochs \citep[aligning with previous studies, e.g.,][]{Stacy11,Riaz+18}. 

(vii) Another important aspect of our model is to explore the merger phenomenon with a perfect inelastic collision model.

(viii) The number of mergers is very sensitive to the fraction of radiation. A lower value of the fraction yields a higher number of mergers. 

(ix) Our numerical model addressed the formation of both low and high-mass protostars, spanning over the range $\approx0.35-75 \ M_{\odot}$. This is due to the implementation of the radiation pressure, that constrain the mass function. 

(x) It is to be noted that while the feedback may not affect the dynamics of the low-mass star due to their low luminosity, it could indirectly do so by affecting the mass of the massive star, and therefore the N-body interactions in the system.

(xi) In addition to the few massive protostars that stay within the cloud, we also observed that around 20\% of the protostars seem to escape the cloud with a mass less than $0.8M_{\odot}$ and could be the candidates for entering the ZAMS stage and subsequently surviving till the present epoch of time.

This work holds significance across various dimensions. First, instead of delving into the complexity of the existing hydrodynamical simulations, we have set up a simplified experiment that can study the evolution of the protostars with significantly lower computational resources. Second, while our model may not exactly align with traditional hydrodynamical simulations, we provide here with a \textit{fast} alternative framework in order to address several issues like N-Body interaction, merging and feedback, which still remains a challenging task in hydrodynamical simulations. Third, his semi-analytical calculation especially implements the merger at a scale of $\approx R_{\odot}$. This cannot be probed in standard hydrodynamical simulations where the resolution is usually not enough to capture the physics in the length scale of the protostellar radii. Fourth, the adaptive time-stepping scheme and the accuracy of this model give a first-hand estimation of the mass function of the protostars. Besides, the existence of both high and low-mass protostars in our simulation also addresses the long-term conjecture on the mass of the protostars. Finally, we wish to conclude that for the first time to the best of our knowledge, we provide here with a model that can study the evolution of the protostars, with some approximations, for a long period of time, which is still not achievable in a 3D high-resolution hydrodynamical simulation. We hope that this model will be of utmost importance for the primordial community.

\begin{acknowledgments}
We thank the Harish-Chandra Research Institute (HRI) for providing a Visiting Fellowship and hospitality to finish this work. We thank Jasjeet Singh Bagla, Athena Stacy, Sharanya Sur for the fruitful discussion. This research has made use of NASA's Astrophysics Data System bibliographic services.
\end{acknowledgments}

\appendix

\section{Tests for the numerical scheme}
\label{sec:test}

We have rigorously tested our numerical calculation with well-established previous studies in order to ensure the correctness and stability of the code. Below, we list the conducted tests.

\subsection{N-Body interaction}
\begin{figure}[h]
\centering
\includegraphics[width=0.3\textwidth]{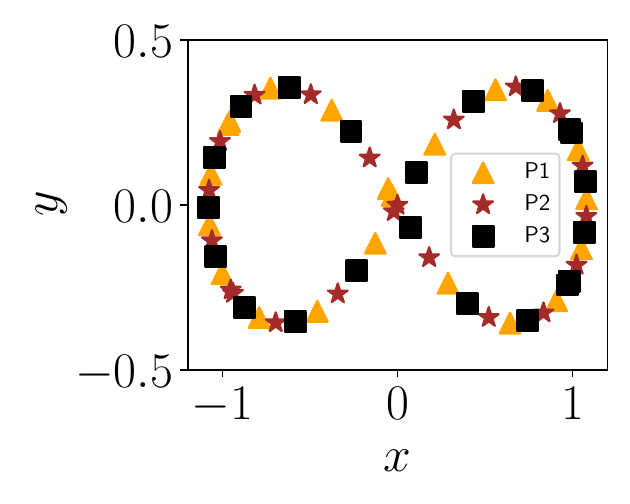}
\caption{The well-established `Figure-8' solution for the three-body problem has been numerically reproduced, ensuring the correctness of our N-Body solver. The trajectories of all three particles P1, P2 and P3 exhibit a similar profile resembling the shape of the number `8'.}
\label{fig:3b}
\end{figure}

The N-Body solver has been verified by reproducing the well-known `Figure-8' solution, which is considered to be a standard solution of the three-body problem \citep{Chenciner2000}. We initialised the three particles with mass $m_1=1,m_2=1,m_3=1$, located at $x_1,y_1=-0.9700436,0.24308753; x_2,y_2=0.0,0.0\ \mathrm{and} \ x_3,y_3=0.9700436,-0.24308753$, with velocity $v_{x1},v_{y1}=0.4662036850,0.4323657300;v_{x2},v_{y2}=-0.93240737,-0.86473146 \ \mathrm{and} \ v_{x3},v_{y3}=0.4662036850,0.4323657300$, respectively, each in dimensionless units. Fig. \ref{fig:3b} reflects the motion of the three particles denoted by the colors `orange', `brown' and `black' in the $x-y$ plane. The trajectories look exactly like the `Figure-8' solution as expected from the solution of three-body problems.

\subsection{Dynamical Evolution}

\begin{figure}[h]
\centering
\includegraphics[width=0.3\textwidth]{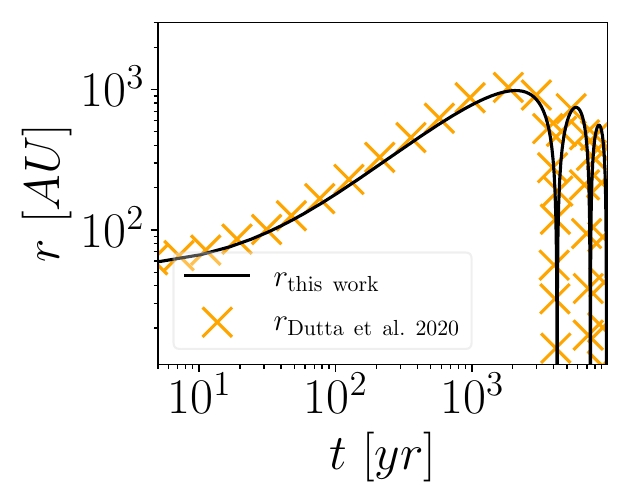}
\caption{The numerical test is performed for a one-particle system in order to verify the correctness of the dynamics. The time evolution of the radial distance (denoted in black) is similar to the previous study by \citet{Dutta+2020}}
\label{fig:acc}
\end{figure}

The accuracy of the dynamical evolution has been rigorously checked with the previous study by \citet{Dutta+2020} that applied to a single accreting protostar. Hence, switching off the gas rotation and feedback process in our current setup corresponds to the result of \citet{Dutta+2020}. The time evolution of the radial distance for a single protostar is depicted in Fig. \ref{fig:acc}, in which, the parameters we initialised with are the following: $m_i=0.03 M_{\odot}$, $r_i=50 \ \mathrm{AU}$, $v_{r,i}=7 \ \mathrm{kms^{-1}}$ and $v_{\phi,i}=0.01 \ \mathrm{kms^{-1}}$.

\subsection{Merging phenomenon}

\begin{figure}[h]
\centering
  \centering
  \includegraphics[width=0.3\textwidth]{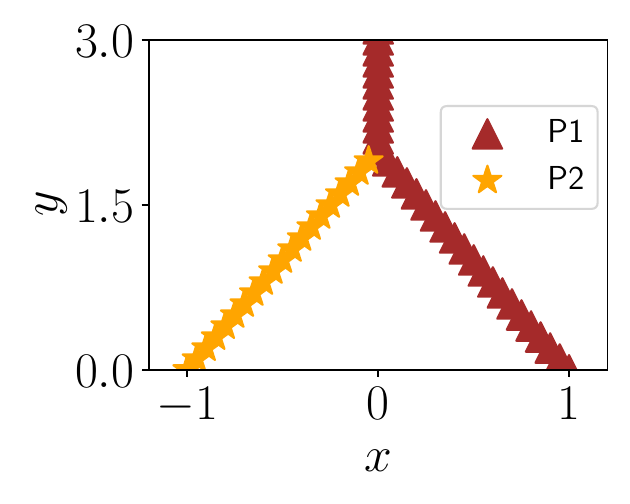}
  \includegraphics[width=0.3\textwidth]{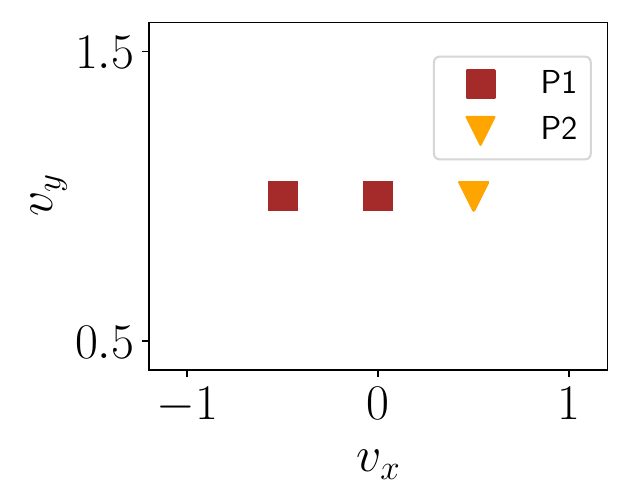} 
  \caption{The trajectories of the two particles, i.e., `P1' (in red) and `P2' (in orange), which merge together, are demonstrated in the $x-y$ plane (left panel). It is to be noted that both the particles are coming from the opposite direction with the same speed, and the particle P2 merges into the particle P1. Hence, the $x$-component of the velocity of the merged P1 becomes 0, whereas the $y$-component remains unchanged.} 
\label{fig:2b}
\end{figure}

The precise implementation of the merging phenomenon has been meticulously testified to the solution of a perfectly inelastic collision between two particles. We initialised the two particles with mass $(m_1,m_2)=(1,1)$, position $(x_1,y_1)=(1,0);(x_2,y_2)=(-1,0)$ and velocity 
$(v_{x,1},v_{y,1})=(-0.5,1.0); (v_{x,2},v_{y,2})=(0.5,1.0)$ respectively, where all variables are dimensionless. The physical characteristics have been clearly demonstrated in the $x-y$ plane in Fig. \ref{fig:2b}. The plot shows that both particles come close to each other, and at some point the second particle merges with the first one. 

After the collision, the $x$-component of the velocity tends to become zero for the merged particle. However, the $y$-component of the velocity remains the same as that of the first particle. We have also verified that the mass of the merged particle equals the combined mass of the first and the second particle. This ensures the implementation of a perfectly inelastic collision. 

\section{Dynamics of the protostars in the five random realizations}\label{sec:diff_ic}

\begin{figure}[h]
   \centering
   \includegraphics[width=0.5\textwidth]{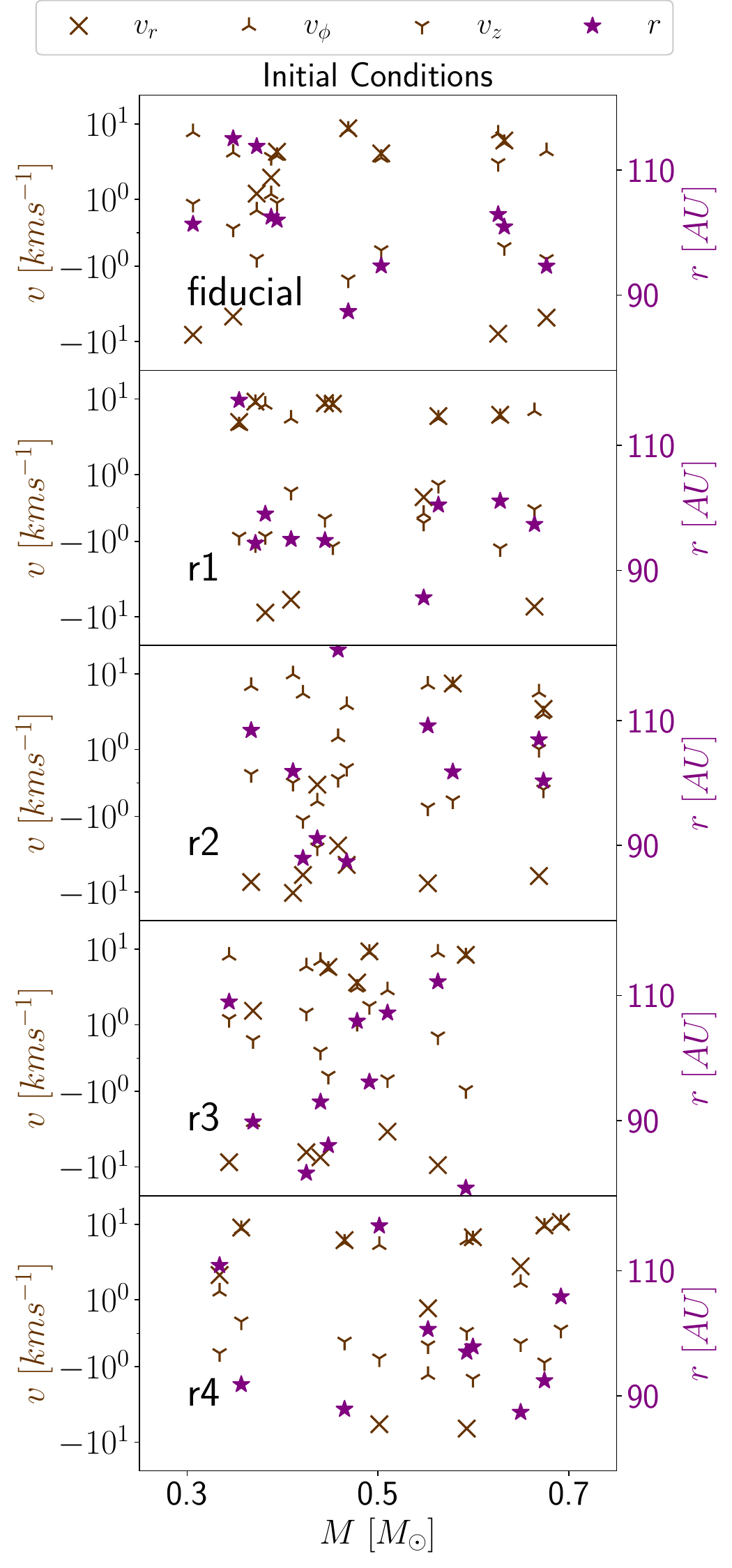} 
   \caption{The \textit{Initial Conditions}, i.e, distribution of initial mass, velocity and radial distance of the ten protostars in the four random realisations, namely r1, r2, r3 and r4 follow the same distribution as the fiducial model. This initial condition corresponds to physical systems explored in previous studies, the details of which are mentioned in \S\ref{sub:ic}.}
   \label{fig:ic_all}
\end{figure}

\begin{figure*}[htp]
    \centering
    \subfigure[Fiducial Model]{
        \includegraphics[width=0.3\textwidth]{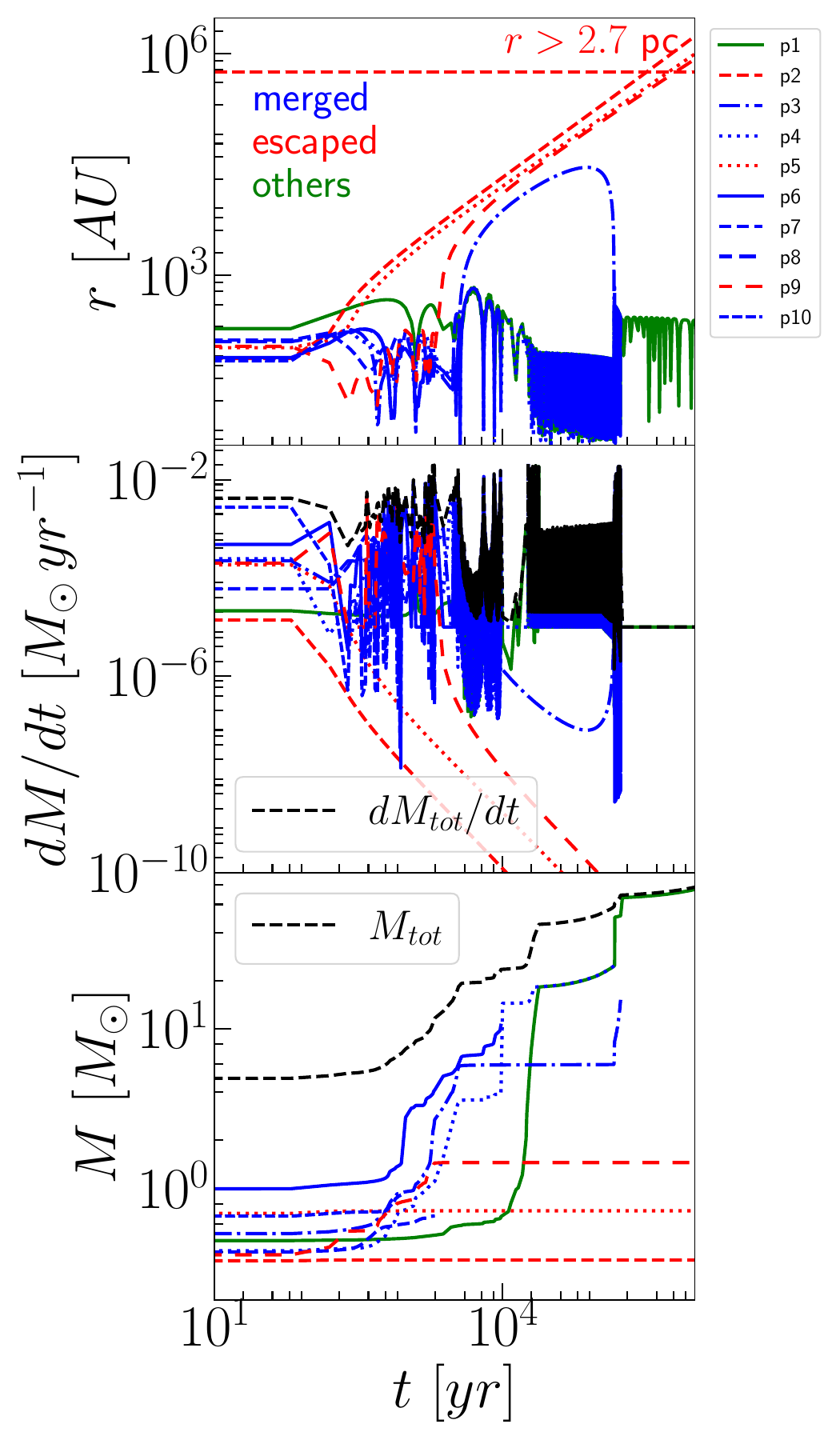}
    }
    \subfigure[Realization `r1']{
        \includegraphics[width=0.3\textwidth]{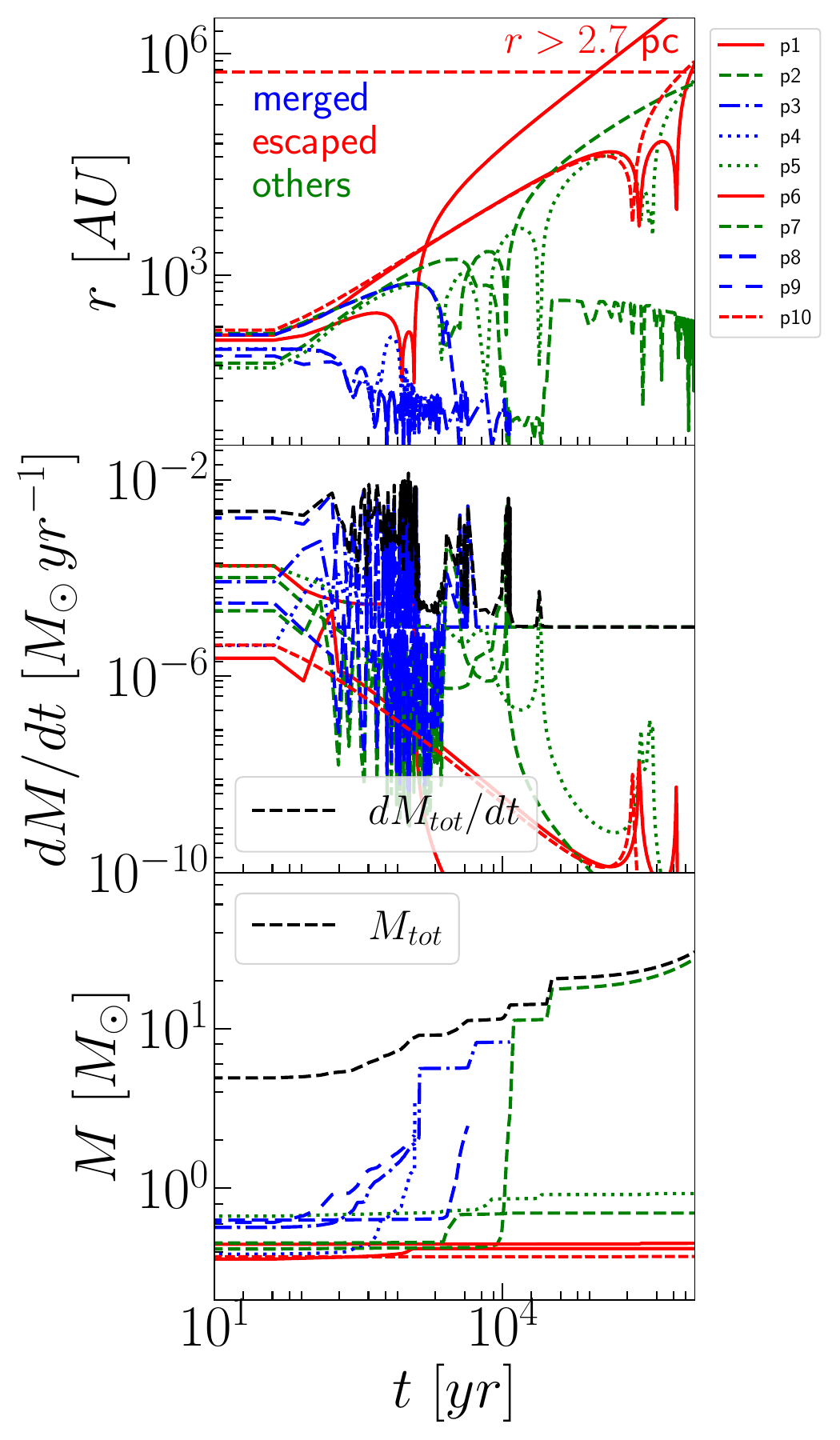}
    }
    \subfigure[Realization `r2']{
        \includegraphics[width=0.3\textwidth]{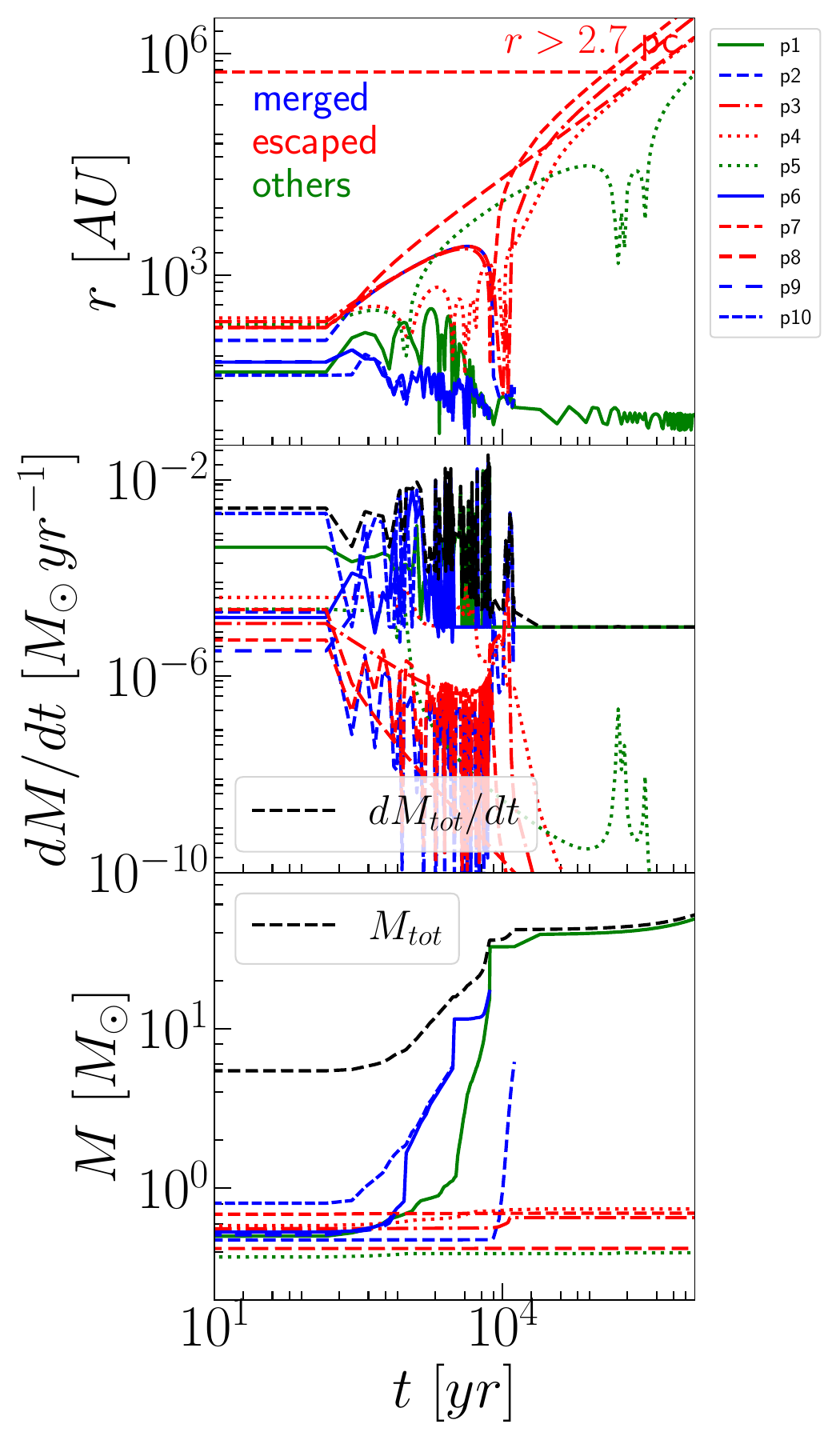}
    }
    \subfigure[Realization `r3']{
        \includegraphics[width=0.3\textwidth]{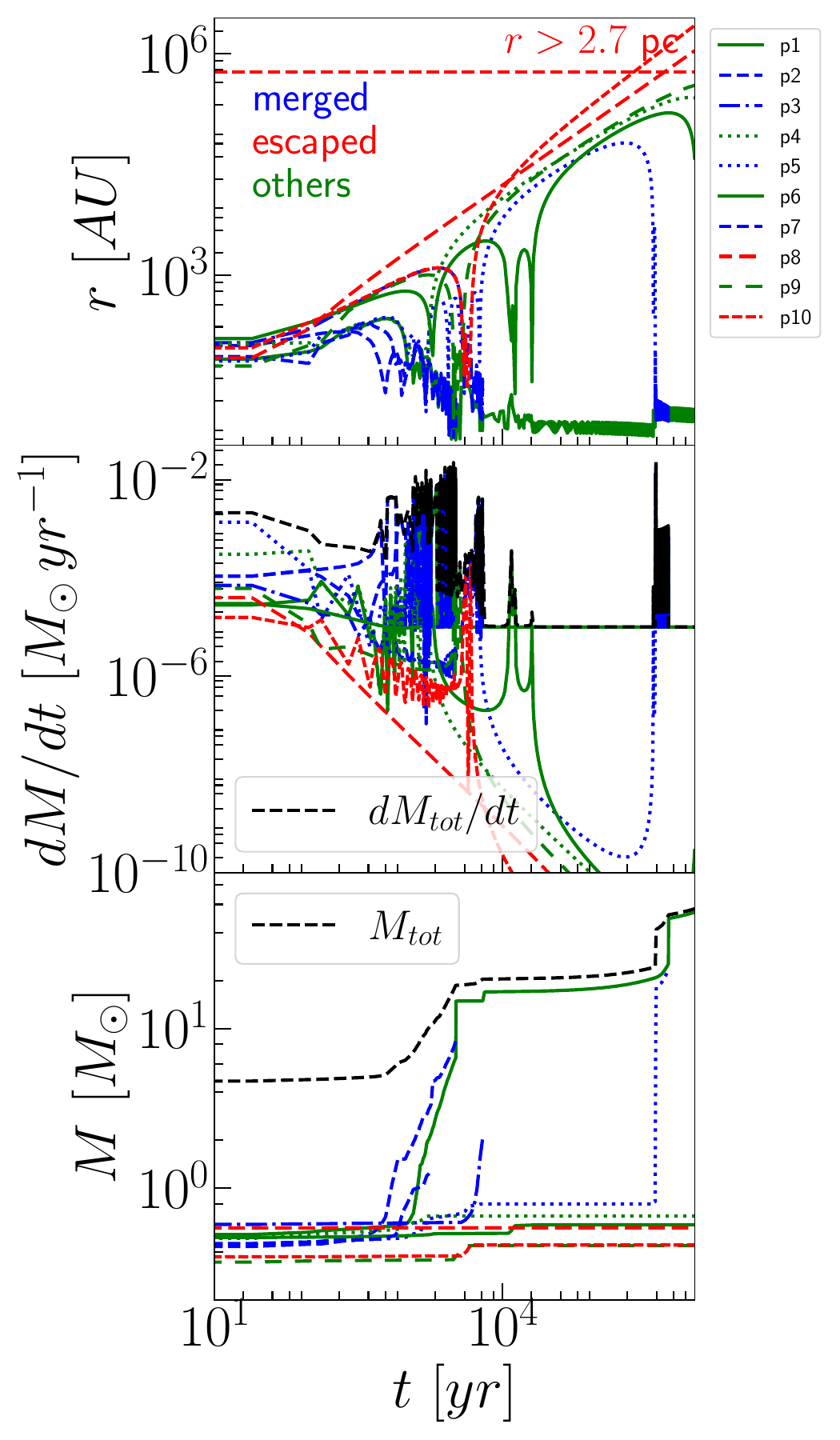}
    }
    \subfigure[Realization `r4']{
        \includegraphics[width=0.3\textwidth]{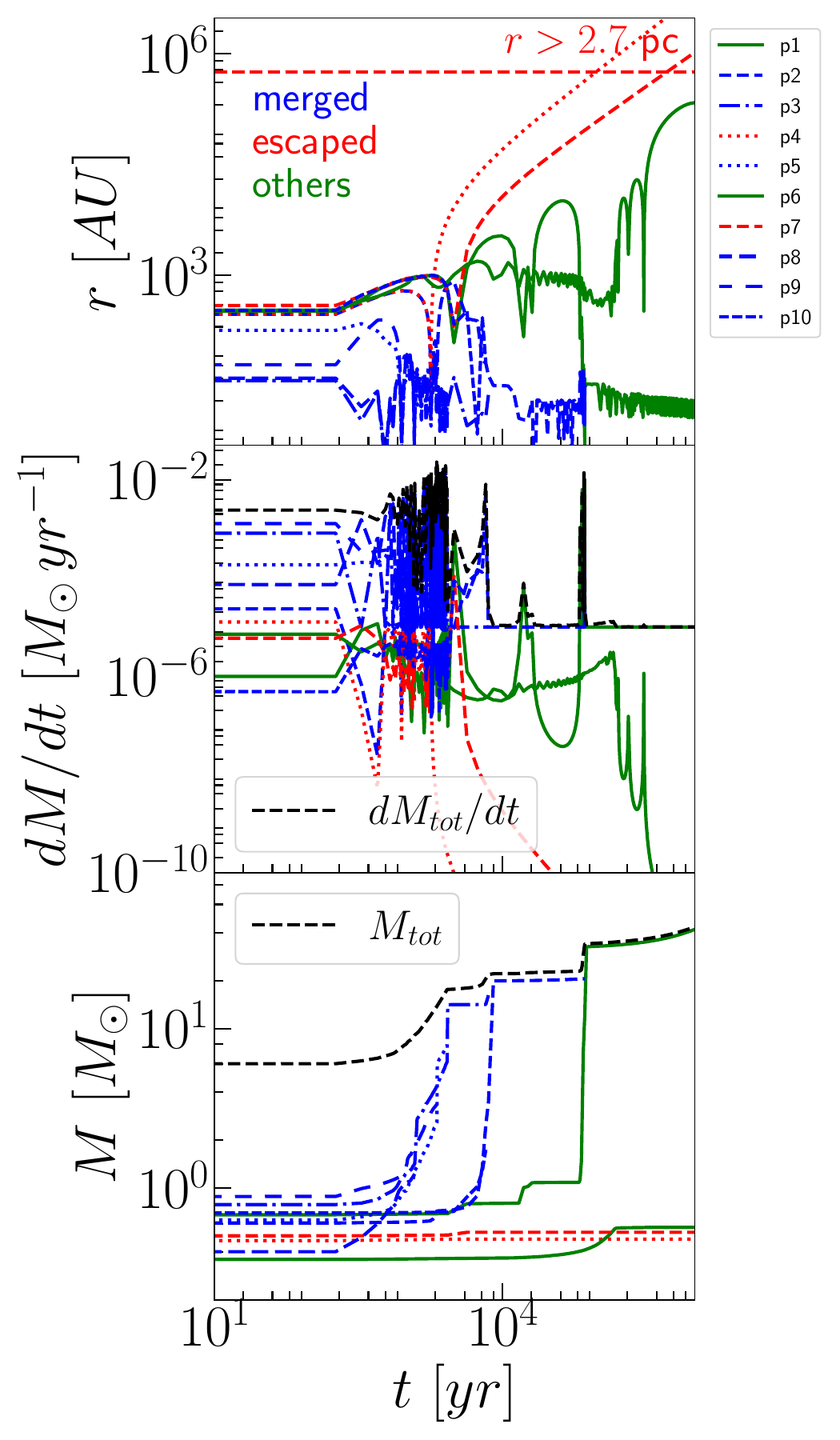}
    }
    \caption{The time evolution of the radial distance (the top panel), accretion rate (the middle panel), and mass (the bottom panel) of the evolving protostar show similar trends in all five realizations. Every realization shows previously observed phenomena such as close encounters, mergers, and the escape phenomenon.}
    \label{fig:all}
\end{figure*}

\begin{figure*}[htp]
    \centering
    \subfigure[Fiducial Model]{
        \includegraphics[width=0.37\textwidth]{Binary.png}
    }
    \subfigure[Realization `r1']{
        \includegraphics[width=0.37\textwidth]{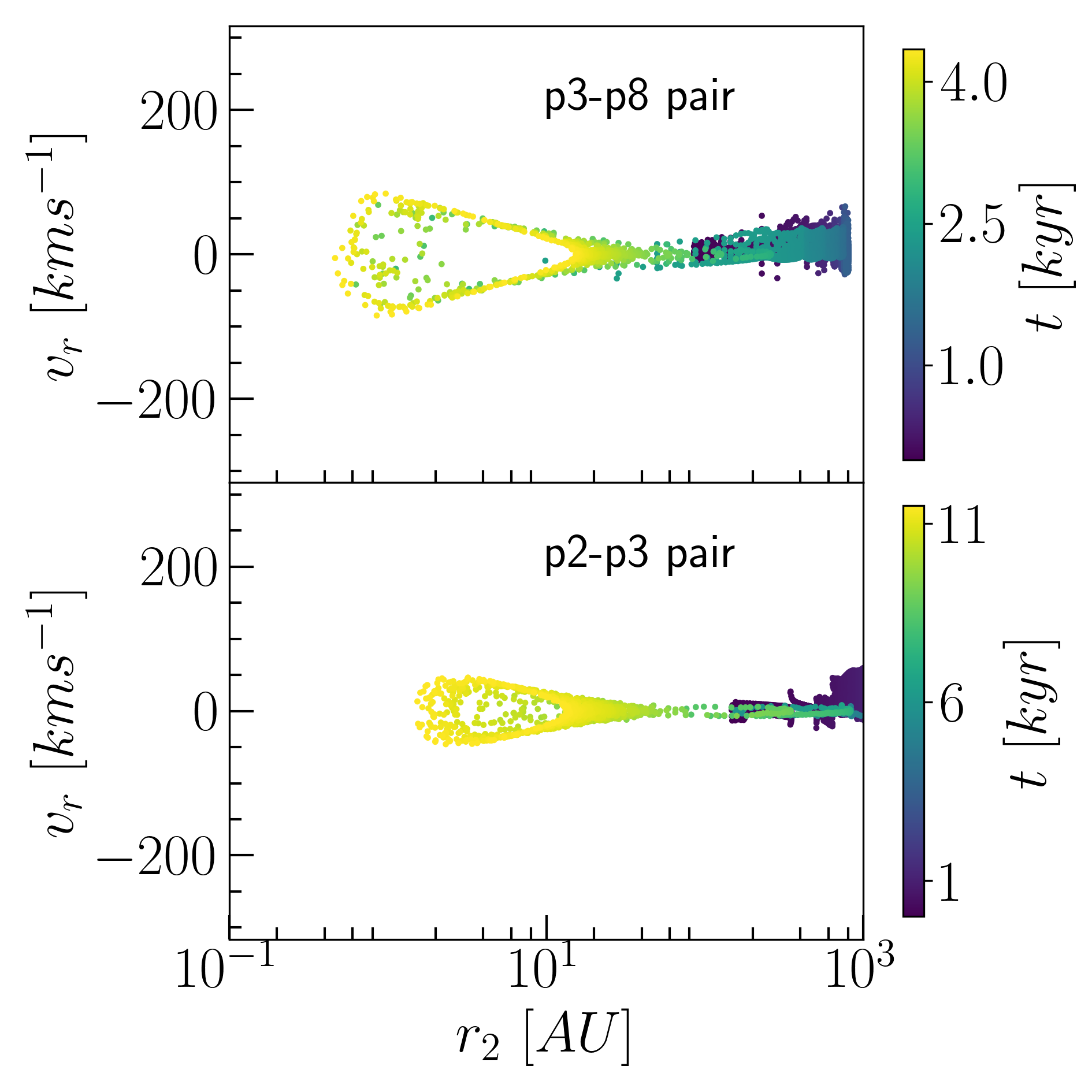}
    }
    \subfigure[Realization `r2']{
        \includegraphics[width=0.37\textwidth]{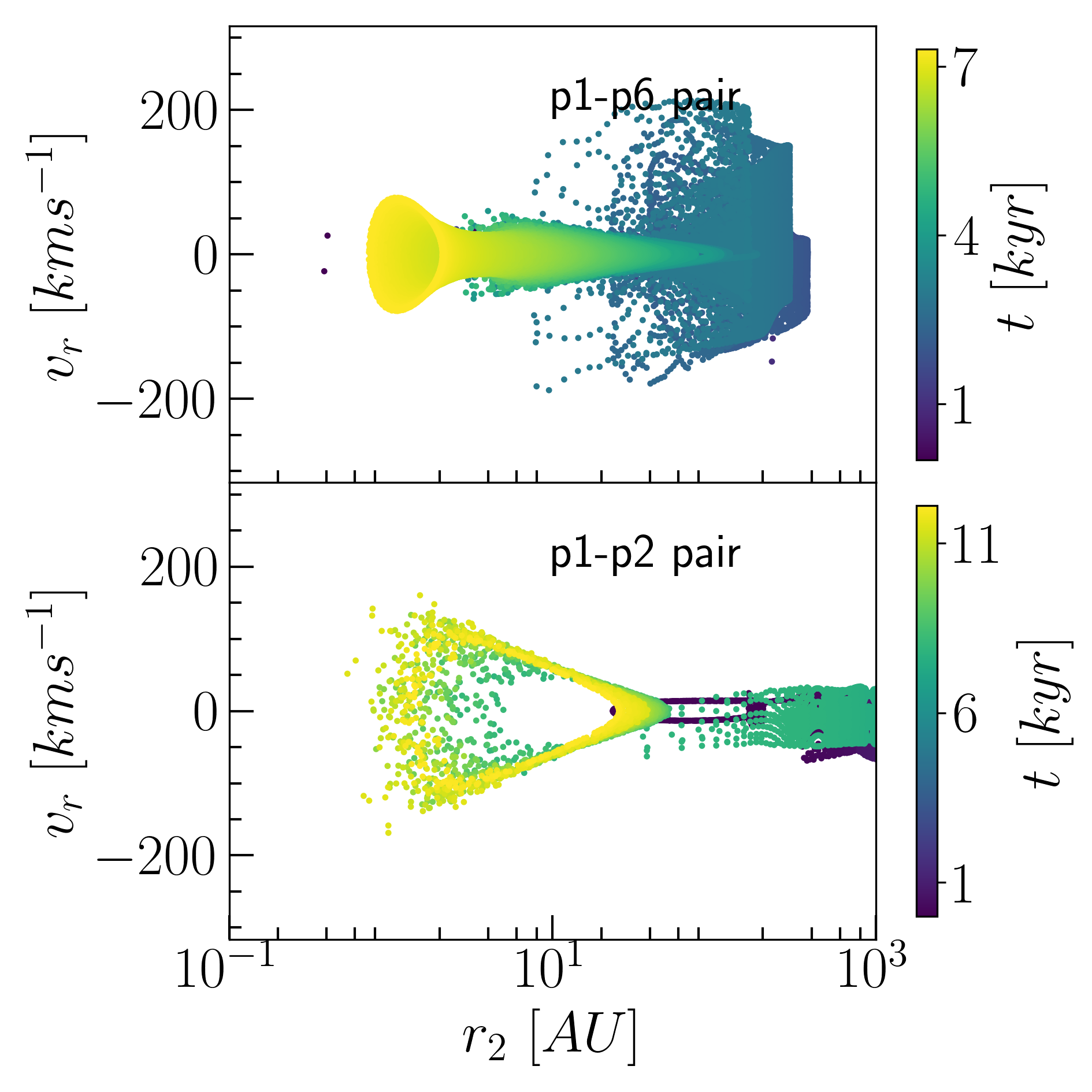}
    }
    \subfigure[Realization `r3']{
        \includegraphics[width=0.37\textwidth]{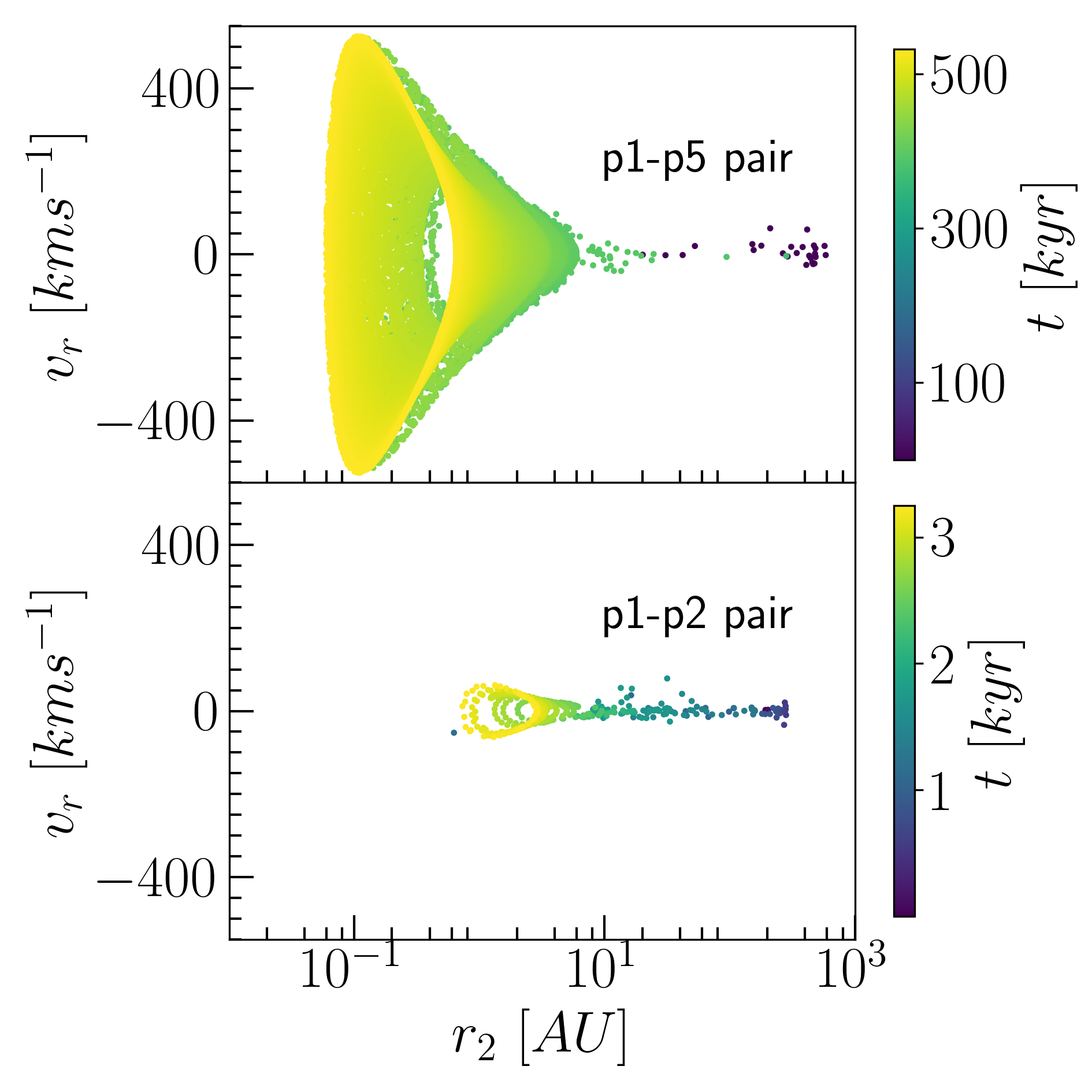}
    }
    \subfigure[Realization `r4']{
        \includegraphics[width=0.37\textwidth]{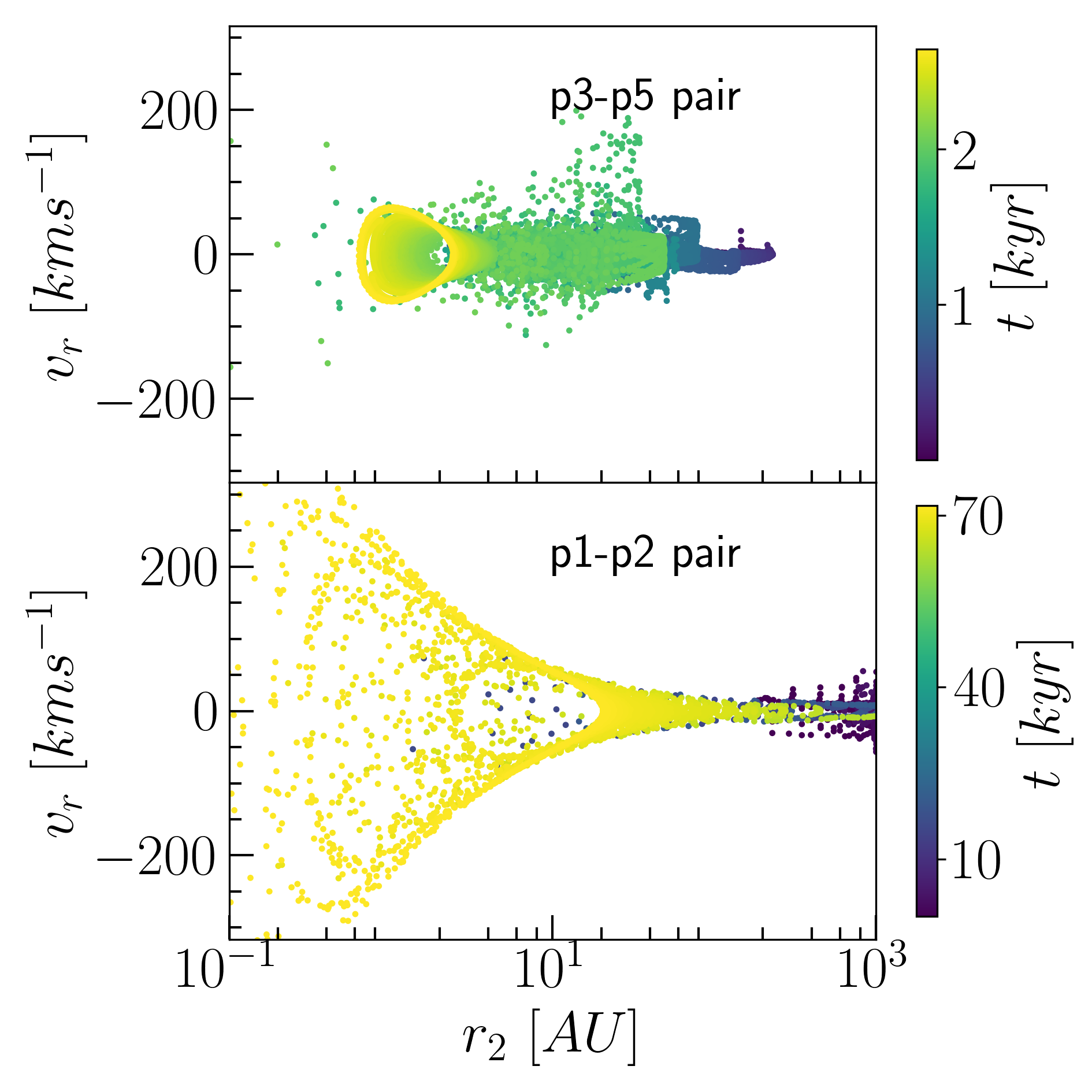}
    }
    \caption{Each realization leads to the formation of binary pairs. Two selected pairs from each realization are illustrated above.}
    \label{fig:binary_all}
\end{figure*}

Fig. \ref{fig:ic_all} shows the initial distribution of the mass, radial distance and the three components of velocity, namely `$v_r$',`$v_\phi$',`$v_z$' of the protostars for the five random realizations. The time evolution of the protostars is demonstrated in Fig. \ref{fig:all}, whereas the formation and evolution of binary pairs are shown in Fig. \ref{fig:binary_all}.

\bibliographystyle{aasjournal}

\begin{thebibliography}{}
\expandafter\ifx\csname natexlab\endcsname\relax\def\natexlab#1{#1}\fi
\providecommand{\url}[1]{\href{#1}{#1}}
\providecommand{\dodoi}[1]{doi:~\href{http://doi.org/#1}{\nolinkurl{#1}}}
\providecommand{\doeprint}[1]{\href{http://ascl.net/#1}{\nolinkurl{http://ascl.net/#1}}}
\providecommand{\doarXiv}[1]{\href{https://arxiv.org/abs/#1}{\nolinkurl{https://arxiv.org/abs/#1}}}

\bibitem[{Abel {et~al.}(2002)Abel, Bryan, \& Norman}]{Abel:2001pr}
Abel, T., Bryan, G.~L., \& Norman, M.~L. 2002, Science, 295, 93,
  \dodoi{10.1126/science.1063991}

\bibitem[{Alister~Seguel {et~al.}(2020)Alister~Seguel, Schleicher, Boekholt,
  Fellhauer, \& Klessen}]{Seguel+2020}
Alister~Seguel, P.~J., Schleicher, D. R.~G., Boekholt, T. C.~N., Fellhauer, M.,
  \& Klessen, R.~S. 2020, Monthly Notices of the Royal Astronomical Society,
  493, 2352, \dodoi{10.1093/mnras/staa456}

\bibitem[{{Banik} \& {Bovy}(2021)}]{Banik+21}
{Banik}, N., \& {Bovy}, J. 2021, \mnras, 504, 648,
  \dodoi{10.1093/mnras/stab886}

\bibitem[{{Barkana}(2018)}]{Barkana+18}
{Barkana}, R. 2018, \nat, 555, 71, \dodoi{10.1038/nature25791}

\bibitem[{{Barkana} \& {Loeb}(2001)}]{Barkana_Loeb_01}
{Barkana}, R., \& {Loeb}, A. 2001, \physrep, 349, 125,
  \dodoi{10.1016/S0370-1573(01)00019-9}

\bibitem[{{Bate} {et~al.}(2003){Bate}, {Bonnell}, \& {Bromm}}]{Bate+03}
{Bate}, M.~R., {Bonnell}, I.~A., \& {Bromm}, V. 2003, \mnras, 339, 577,
  \dodoi{10.1046/j.1365-8711.2003.06210.x}

\bibitem[{{Bate} {et~al.}(1995){Bate}, {Bonnell}, \& {Price}}]{Bate+95}
{Bate}, M.~R., {Bonnell}, I.~A., \& {Price}, N.~M. 1995, \mnras, 277, 362,
  \dodoi{10.1093/mnras/277.2.362}

\bibitem[{{Baumgardt} \& {Makino}(2003)}]{Holger+03}
{Baumgardt}, H., \& {Makino}, J. 2003, \mnras, 340, 227,
  \dodoi{10.1046/j.1365-8711.2003.06286.x}

\bibitem[{{Becerra} {et~al.}(2015){Becerra}, {Greif}, {Springel}, \&
  {Hernquist}}]{Becerra_15}
{Becerra}, F., {Greif}, T.~H., {Springel}, V., \& {Hernquist}, L.~E. 2015,
  \mnras, 446, 2380, \dodoi{10.1093/mnras/stu2284}

\bibitem[{{Bovino} {et~al.}(2014){Bovino}, {Schleicher}, \&
  {Grassi}}]{Bovino+14}
{Bovino}, S., {Schleicher}, D.~R.~G., \& {Grassi}, T. 2014, \aap, 561, A13,
  \dodoi{10.1051/0004-6361/201322387}

\bibitem[{Bromm \& Larson(2004)}]{Bromm+04}
Bromm, V., \& Larson, R.~B. 2004, Annual Review of Astronomy and Astrophysics,
  42, 79, \dodoi{10.1146/annurev.astro.42.053102.134034}

\bibitem[{{Caffau} {et~al.}(2011){Caffau}, {Bonifacio}, {Fran{\c{c}}ois},
  {Sbordone}, {Monaco}, {Spite}, {Spite}, {Ludwig}, {Cayrel}, {Zaggia},
  {Hammer}, {Randich}, {Molaro}, \& {Hill}}]{caffau+11}
{Caffau}, E., {Bonifacio}, P., {Fran{\c{c}}ois}, P., {et~al.} 2011, \nat, 477,
  67, \dodoi{10.1038/nature10377}

\bibitem[{Chambers \& Wetherill(1998)}]{CHAMBERS98}
Chambers, J., \& Wetherill, G. 1998, Icarus, 136, 304,
  \dodoi{https://doi.org/10.1006/icar.1998.6007}

\bibitem[{Chen {et~al.}(2014)Chen, Heger, Woosley, Almgren, \&
  Whalen}]{Chen_2014}
Chen, K.-J., Heger, A., Woosley, S., Almgren, A., \& Whalen, D.~J. 2014, The
  Astrophysical Journal, 792, 44, \dodoi{10.1088/0004-637X/792/1/44}

\bibitem[{Chenciner \& Montgomery(2000)}]{Chenciner2000}
Chenciner, A., \& Montgomery, R. 2000, Annals of Mathematics. Second Series,
  152, 881.
\newblock \url{http://eudml.org/doc/121861}

\bibitem[{{Chiaki} \& {Yoshida}(2022)}]{Chiaki+22}
{Chiaki}, G., \& {Yoshida}, N. 2022, \mnras, 510, 5199,
  \dodoi{10.1093/mnras/stab2799}

\bibitem[{{Choudhuri}(2010)}]{ARC10}
{Choudhuri}, A.~R. 2010, {Astrophysics for Physicists}

\bibitem[{{Ciardi} \& {Ferrara}(2005)}]{Ciardi_Ferrara_05}
{Ciardi}, B., \& {Ferrara}, A. 2005, \ssr, 116, 625,
  \dodoi{10.1007/s11214-005-3592-0}

\bibitem[{{Clark} {et~al.}(2011){Clark}, {Glover}, {Smith}, {Greif}, {Klessen},
  \& {Bromm}}]{Clark+11}
{Clark}, P.~C., {Glover}, S. C.~O., {Smith}, R.~J., {et~al.} 2011, Science,
  331, 1040, \dodoi{10.1126/science.1198027}

\bibitem[{Dutta(2015)}]{Dutta_2015}
Dutta, J. 2015, The Astrophysical Journal, 811, 98,
  \dodoi{10.1088/0004-637X/811/2/98}

\bibitem[{{Dutta}(2016{\natexlab{a}})}]{Dutta16}
{Dutta}, J. 2016{\natexlab{a}}, \aap, 585, A59,
  \dodoi{10.1051/0004-6361/201526747}

\bibitem[{{Dutta}(2016{\natexlab{b}})}]{Dutta:2015tcl}
---. 2016{\natexlab{b}}, \apss, 361, 35, \dodoi{10.1007/s10509-015-2622-y}

\bibitem[{{Dutta} {et~al.}(2015){Dutta}, {Nath}, {Clark}, \&
  {Klessen}}]{Dutta_15_3}
{Dutta}, J., {Nath}, B.~B., {Clark}, P.~C., \& {Klessen}, R.~S. 2015, \mnras,
  450, 202, \dodoi{10.1093/mnras/stv664}

\bibitem[{Dutta {et~al.}(2020)Dutta, Sur, Stacy, \& Bagla}]{Dutta+2020}
Dutta, J., Sur, S., Stacy, A., \& Bagla, J.~S. 2020, The Astrophysical Journal,
  901, 16, \dodoi{10.3847/1538-4357/abadf8}

\bibitem[{Edgar(2004)}]{EDGAR04}
Edgar, R. 2004, New Astronomy Reviews, 48, 843,
  \dodoi{https://doi.org/10.1016/j.newar.2004.06.001}

\bibitem[{{Frebel} \& {Norris}(2015)}]{Frebel_Norris_15}
{Frebel}, A., \& {Norris}, J.~E. 2015, \araa, 53, 631,
  \dodoi{10.1146/annurev-astro-082214-122423}

\bibitem[{Glover \& Abel(2008)}]{Glover_Abel_08}
Glover, S. C.~O., \& Abel, T. 2008, Monthly Notices of the Royal Astronomical
  Society, 388, 1627, \dodoi{10.1111/j.1365-2966.2008.13224.x}

\bibitem[{Greif {et~al.}(2012)Greif, Bromm, Clark, Glover, Smith, Klessen,
  Yoshida, \& Springel}]{Greif+12}
Greif, T.~H., Bromm, V., Clark, P.~C., {et~al.} 2012, Monthly Notices of the
  Royal Astronomical Society, 424, 399,
  \dodoi{10.1111/j.1365-2966.2012.21212.x}

\bibitem[{Greif {et~al.}(2011)Greif, Springel, White, Glover, Clark, Smith,
  Klessen, \& Bromm}]{Greif_2011}
Greif, T.~H., Springel, V., White, S. D.~M., {et~al.} 2011, The Astrophysical
  Journal, 737, 75, \dodoi{10.1088/0004-637X/737/2/75}

\bibitem[{{Haemmerl{\'e}} {et~al.}(2020){Haemmerl{\'e}}, {Mayer}, {Klessen},
  {Hosokawa}, {Madau}, \& {Bromm}}]{Haemmerle+20}
{Haemmerl{\'e}}, L., {Mayer}, L., {Klessen}, R.~S., {et~al.} 2020, \ssr, 216,
  48, \dodoi{10.1007/s11214-020-00673-y}

\bibitem[{{Haemmerl{\'e}} {et~al.}(2018){Haemmerl{\'e}}, {Woods}, {Klessen},
  {Heger}, \& {Whalen}}]{Haemmerle+18}
{Haemmerl{\'e}}, L., {Woods}, T.~E., {Klessen}, R.~S., {Heger}, A., \&
  {Whalen}, D.~J. 2018, \mnras, 474, 2757, \dodoi{10.1093/mnras/stx2919}

\bibitem[{Haemmerlé {et~al.}(2017)Haemmerlé, Woods, Klessen, Heger, \&
  Whalen}]{Haemmerle+17}
Haemmerlé, L., Woods, T.~E., Klessen, R.~S., Heger, A., \& Whalen, D.~J. 2017,
  Monthly Notices of the Royal Astronomical Society, 474, 2757,
  \dodoi{10.1093/mnras/stx2919}

\bibitem[{{Haiman}(2011)}]{Haiman_2011}
{Haiman}, Z. 2011, \nat, 472, 47, \dodoi{10.1038/472047a}

\bibitem[{{Hartwig} {et~al.}(2015){Hartwig}, {Bromm}, {Klessen}, \&
  {Glover}}]{Hartwig+15}
{Hartwig}, T., {Bromm}, V., {Klessen}, R.~S., \& {Glover}, S. C.~O. 2015,
  \mnras, 447, 3892, \dodoi{10.1093/mnras/stu2740}

\bibitem[{Hirano {et~al.}(2015)Hirano, Hosokawa, Yoshida, Omukai, \&
  Yorke}]{Hirano+15}
Hirano, S., Hosokawa, T., Yoshida, N., Omukai, K., \& Yorke, H.~W. 2015,
  Monthly Notices of the Royal Astronomical Society, 448, 568,
  \dodoi{10.1093/mnras/stv044}

\bibitem[{Hosokawa {et~al.}(2011)Hosokawa, Omukai, Yoshida, \&
  Yorke}]{Hosokawa+11}
Hosokawa, T., Omukai, K., Yoshida, N., \& Yorke, H.~W. 2011, Science, 334,
  1250, \dodoi{10.1126/science.1207433}

\bibitem[{{Inoue} \& {Yoshida}(2020)}]{Inoue+20}
{Inoue}, S., \& {Yoshida}, N. 2020, \mnras, 491, L24,
  \dodoi{10.1093/mnrasl/slz160}

\bibitem[{Ishigaki {et~al.}(2018)Ishigaki, Tominaga, Kobayashi, \&
  Nomoto}]{Ishigaki_2018}
Ishigaki, M.~N., Tominaga, N., Kobayashi, C., \& Nomoto, K. 2018, The
  Astrophysical Journal, 857, 46, \dodoi{10.3847/1538-4357/aab3de}

\bibitem[{Ishiyama {et~al.}(2016)Ishiyama, Sudo, Yokoi, Hasegawa, Tominaga, \&
  Susa}]{Ishiyama_2016}
Ishiyama, T., Sudo, K., Yokoi, S., {et~al.} 2016, The Astrophysical Journal,
  826, 9, \dodoi{10.3847/0004-637X/826/1/9}

\bibitem[{{Jaura} {et~al.}(2018){Jaura}, {Glover}, {Klessen}, \&
  {Paardekooper}}]{Jaura+18}
{Jaura}, O., {Glover}, S.~C.~O., {Klessen}, R.~S., \& {Paardekooper}, J.~P.
  2018, \mnras, 475, 2822, \dodoi{10.1093/mnras/stx3356}

\bibitem[{{Jaura} {et~al.}(2022){Jaura}, {Glover}, {Wollenberg}, {Klessen},
  {Geen}, \& {Haemmerl{\'e}}}]{Jaura+22}
{Jaura}, O., {Glover}, S. C.~O., {Wollenberg}, K. M.~J., {et~al.} 2022, \mnras,
  512, 116, \dodoi{10.1093/mnras/stac487}

\bibitem[{Kirihara {et~al.}(2019)Kirihara, Tanikawa, \& Ishiyama}]{Kirihara+19}
Kirihara, T., Tanikawa, A., \& Ishiyama, T. 2019, Monthly Notices of the Royal
  Astronomical Society, 486, 5917, \dodoi{10.1093/mnras/stz1277}

\bibitem[{Klessen \& Glover(2023)}]{Klessen:2023qmc}
Klessen, R.~S., \& Glover, S.~C. 2023, Annual Review of Astronomy and
  Astrophysics, 61, 65, \dodoi{10.1146/annurev-astro-071221-053453}

\bibitem[{{Komiya} {et~al.}(2015){Komiya}, {Suda}, \& {Fujimoto}}]{Komiya+15}
{Komiya}, Y., {Suda}, T., \& {Fujimoto}, M.~Y. 2015, \apjl, 808, L47,
  \dodoi{10.1088/2041-8205/808/2/L47}

\bibitem[{{Komiya} {et~al.}(2016){Komiya}, {Suda}, \& {Fujimoto}}]{Komiya+16}
---. 2016, \apj, 820, 59, \dodoi{10.3847/0004-637X/820/1/59}

\bibitem[{Krumholz {et~al.}(2009)Krumholz, Klein, McKee, Offner, \&
  Cunningham}]{Krumholz+09}
Krumholz, M.~R., Klein, R.~I., McKee, C.~F., Offner, S. S.~R., \& Cunningham,
  A.~J. 2009, Science, 323, 754, \dodoi{10.1126/science.1165857}

\bibitem[{{Kuiper} {et~al.}(2010){Kuiper}, {Klahr}, {Beuther}, \&
  {Henning}}]{Kuiper+10}
{Kuiper}, R., {Klahr}, H., {Beuther}, H., \& {Henning}, T. 2010, \apj, 722,
  1556, \dodoi{10.1088/0004-637X/722/2/1556}

\bibitem[{{Kulkarni} {et~al.}(2019){Kulkarni}, {Visbal}, \&
  {Bryan}}]{Kulkarni+19}
{Kulkarni}, M., {Visbal}, E., \& {Bryan}, G.~L. 2019, \apj, 882, 178,
  \dodoi{10.3847/1538-4357/ab35e2}

\bibitem[{{Lambrechts} {et~al.}(2019){Lambrechts}, {Morbidelli}, {Jacobson},
  {Johansen}, {Bitsch}, {Izidoro}, \& {Raymond}}]{Michiel+19}
{Lambrechts}, M., {Morbidelli}, A., {Jacobson}, S.~A., {et~al.} 2019, \aap,
  627, A83, \dodoi{10.1051/0004-6361/201834229}

\bibitem[{{Larson}(1969)}]{Larson+69}
{Larson}, R.~B. 1969, \mnras, 145, 271, \dodoi{10.1093/mnras/145.3.271}

\bibitem[{Latif {et~al.}(2022)Latif, Whalen, \& Khochfar}]{Latif_2022}
Latif, M.~A., Whalen, D., \& Khochfar, S. 2022, The Astrophysical Journal, 925,
  28, \dodoi{10.3847/1538-4357/ac3916}

\bibitem[{Li {et~al.}(2020)Li, Lai, Anderson, \& Pu}]{Li+20}
Li, J., Lai, D., Anderson, K.~R., \& Pu, B. 2020, Monthly Notices of the Royal
  Astronomical Society, 501, 1621, \dodoi{10.1093/mnras/staa3779}

\bibitem[{{Machida} \& {Doi}(2013)}]{Machida+13}
{Machida}, M.~N., \& {Doi}, K. 2013, \mnras, 435, 3283,
  \dodoi{10.1093/mnras/stt1524}

\bibitem[{{Machida} {et~al.}(2008){Machida}, {Omukai}, {Matsumoto}, \&
  {Inutsuka}}]{Machida+08}
{Machida}, M.~N., {Omukai}, K., {Matsumoto}, T., \& {Inutsuka}, S.-i. 2008,
  \apj, 677, 813, \dodoi{10.1086/533434}

\bibitem[{{Marigo} {et~al.}(2001){Marigo}, {Girardi}, {Chiosi}, \&
  {Wood}}]{Marigo+01}
{Marigo}, P., {Girardi}, L., {Chiosi}, C., \& {Wood}, P.~R. 2001, \aap, 371,
  152, \dodoi{10.1051/0004-6361:20010309}

\bibitem[{Matsumoto {et~al.}(1997)Matsumoto, Hanawa, \&
  Nakamura}]{Matsumoto_1997}
Matsumoto, T., Hanawa, T., \& Nakamura, F. 1997, The Astrophysical Journal,
  478, 569, \dodoi{10.1086/303822}

\bibitem[{{Navarro} {et~al.}(1996){Navarro}, {Frenk}, \& {White}}]{Navarro+96}
{Navarro}, J.~F., {Frenk}, C.~S., \& {White}, S. D.~M. 1996, \apj, 462, 563,
  \dodoi{10.1086/177173}

\bibitem[{{Ogihara} \& {Ida}(2009)}]{Ogihara+09}
{Ogihara}, M., \& {Ida}, S. 2009, \apj, 699, 824,
  \dodoi{10.1088/0004-637X/699/1/824}

\bibitem[{{Omukai} \& {Nishi}(1998)}]{Omukai+98}
{Omukai}, K., \& {Nishi}, R. 1998, \apj, 508, 141, \dodoi{10.1086/306395}

\bibitem[{Omukai \& Palla(2003)}]{Omukai_2003}
Omukai, K., \& Palla, F. 2003, The Astrophysical Journal, 589, 677,
  \dodoi{10.1086/374810}

\bibitem[{{Padmanabhan} \& {Loeb}(2022)}]{Padmanabhan+22}
{Padmanabhan}, H., \& {Loeb}, A. 2022, General Relativity and Gravitation, 54,
  24, \dodoi{10.1007/s10714-022-02909-4}

\bibitem[{{Palla} {et~al.}(1983){Palla}, {Salpeter}, \& {Stahler}}]{Palla_1983}
{Palla}, F., {Salpeter}, E.~E., \& {Stahler}, S.~W. 1983, \apj, 271, 632,
  \dodoi{10.1086/161231}

\bibitem[{{Placco} {et~al.}(2015){Placco}, {Frebel}, {Lee}, {Jacobson},
  {Beers}, {Pena}, {Chan}, \& {Heger}}]{Placco+15}
{Placco}, V.~M., {Frebel}, A., {Lee}, Y.~S., {et~al.} 2015, \apj, 809, 136,
  \dodoi{10.1088/0004-637X/809/2/136}

\bibitem[{{Portegies Zwart} \& {McMillan}(2002)}]{Zwart+02}
{Portegies Zwart}, S.~F., \& {McMillan}, S. L.~W. 2002, \apj, 576, 899,
  \dodoi{10.1086/341798}

\bibitem[{Prole {et~al.}(2021)Prole, Clark, Klessen, \& Glover}]{Prole_21}
Prole, L.~R., Clark, P.~C., Klessen, R.~S., \& Glover, S. C.~O. 2021, Monthly
  Notices of the Royal Astronomical Society, 510, 4019,
  \dodoi{10.1093/mnras/stab3697}

\bibitem[{Raghuvanshi \& Dutta(2023)}]{Raghuvanshi+23}
Raghuvanshi, S.~P., \& Dutta, J. 2023, The Astrophysical Journal, 944, 76,
  \dodoi{10.3847/1538-4357/acac30}

\bibitem[{{Riaz} {et~al.}(2018){Riaz}, {Bovino}, {Vanaverbeke}, \&
  {Schleicher}}]{Riaz+18}
{Riaz}, R., {Bovino}, S., {Vanaverbeke}, S., \& {Schleicher}, D.~R.~G. 2018,
  \mnras, 479, 667, \dodoi{10.1093/mnras/sty1635}

\bibitem[{{Rybicki} \& {Lightman}(1986)}]{Rybicki+86}
{Rybicki}, G.~B., \& {Lightman}, A.~P. 1986, {Radiative Processes in
  Astrophysics}

\bibitem[{{Safarzadeh} \& {Haiman}(2020)}]{Safarzadeh+2020}
{Safarzadeh}, M., \& {Haiman}, Z. 2020, \apjl, 903, L21,
  \dodoi{10.3847/2041-8213/abc253}

\bibitem[{Sales {et~al.}(2014)Sales, Marinacci, Springel, \&
  Petkova}]{Sales+14}
Sales, L.~V., Marinacci, F., Springel, V., \& Petkova, M. 2014, Monthly Notices
  of the Royal Astronomical Society, 439, 2990, \dodoi{10.1093/mnras/stu155}

\bibitem[{{Salvadori} {et~al.}(2010){Salvadori}, {Ferrara}, {Schneider},
  {Scannapieco}, \& {Kawata}}]{Salvadori+10}
{Salvadori}, S., {Ferrara}, A., {Schneider}, R., {Scannapieco}, E., \&
  {Kawata}, D. 2010, \mnras, 401, L5, \dodoi{10.1111/j.1745-3933.2009.00772.x}

\bibitem[{Santoliquido {et~al.}(2023)Santoliquido, Mapelli, Iorio, Costa,
  Glover, Hartwig, Klessen, \& Merli}]{Santoliquido_23}
Santoliquido, F., Mapelli, M., Iorio, G., {et~al.} 2023, Monthly Notices of the
  Royal Astronomical Society, 524, 307, \dodoi{10.1093/mnras/stad1860}

\bibitem[{{Schauer} {et~al.}(2022){Schauer}, {Bromm}, {Drory}, \&
  {Boylan-Kolchin}}]{Schauer+22}
{Schauer}, A. T.~P., {Bromm}, V., {Drory}, N., \& {Boylan-Kolchin}, M. 2022,
  \apjl, 934, L6, \dodoi{10.3847/2041-8213/ac7f9a}

\bibitem[{{Sharda} {et~al.}(2019){Sharda}, {Krumholz}, \&
  {Federrath}}]{Sharda+19}
{Sharda}, P., {Krumholz}, M.~R., \& {Federrath}, C. 2019, \mnras, 490, 513,
  \dodoi{10.1093/mnras/stz2618}

\bibitem[{{Smith} {et~al.}(2018){Smith}, {Regan}, {Downes}, {Norman}, {O'Shea},
  \& {Wise}}]{Smith_2018}
{Smith}, B.~D., {Regan}, J.~A., {Downes}, T.~P., {et~al.} 2018, \mnras, 480,
  3762, \dodoi{10.1093/mnras/sty2103}

\bibitem[{Smith {et~al.}(2011)Smith, Glover, Clark, Greif, \&
  Klessen}]{Smith+11}
Smith, R.~J., Glover, S. C.~O., Clark, P.~C., Greif, T., \& Klessen, R.~S.
  2011, Monthly Notices of the Royal Astronomical Society, 414, 3633,
  \dodoi{10.1111/j.1365-2966.2011.18659.x}

\bibitem[{{Springel} {et~al.}(2021){Springel}, {Pakmor}, {Zier}, \&
  {Reinecke}}]{Springel+21}
{Springel}, V., {Pakmor}, R., {Zier}, O., \& {Reinecke}, M. 2021, \mnras, 506,
  2871, \dodoi{10.1093/mnras/stab1855}

\bibitem[{{Stacy}(2011)}]{Stacy11}
{Stacy}, A. 2011, in American Astronomical Society Meeting Abstracts, Vol. 217,
  American Astronomical Society Meeting Abstracts \#217, 133.02

\bibitem[{Stacy {et~al.}(2016)Stacy, Bromm, \& Lee}]{Stacy+16}
Stacy, A., Bromm, V., \& Lee, A.~T. 2016, Monthly Notices of the Royal
  Astronomical Society, 462, 1307, \dodoi{10.1093/mnras/stw1728}

\bibitem[{Stacy {et~al.}(2010)Stacy, Greif, \& Bromm}]{Stacy+10}
Stacy, A., Greif, T.~H., \& Bromm, V. 2010, Monthly Notices of the Royal
  Astronomical Society, 403, 45, \dodoi{10.1111/j.1365-2966.2009.16113.x}

\bibitem[{Stacy {et~al.}(2012)Stacy, Greif, \& Bromm}]{Stacy+12}
---. 2012, Monthly Notices of the Royal Astronomical Society, 422, 290,
  \dodoi{10.1111/j.1365-2966.2012.20605.x}

\bibitem[{{Starkenburg} {et~al.}(2017){Starkenburg}, {Martin}, {Youakim},
  {Aguado}, {Allende Prieto}, {Arentsen}, {Bernard}, {Bonifacio}, {Caffau},
  {Carlberg}, {C{\^o}t{\'e}}, {Fouesneau}, {Fran{\c{c}}ois}, {Franke},
  {Gonz{\'a}lez Hern{\'a}ndez}, {Gwyn}, {Hill}, {Ibata}, {Jablonka},
  {Longeard}, {McConnachie}, {Navarro}, {S{\'a}nchez-Janssen}, {Tolstoy}, \&
  {Venn}}]{Else+17}
{Starkenburg}, E., {Martin}, N., {Youakim}, K., {et~al.} 2017, \mnras, 471,
  2587, \dodoi{10.1093/mnras/stx1068}

\bibitem[{{Sterzik} {et~al.}(2003){Sterzik}, {Durisen}, \&
  {Zinnecker}}]{Sterzik+03}
{Sterzik}, M.~F., {Durisen}, R.~H., \& {Zinnecker}, H. 2003, \aap, 411, 91,
  \dodoi{10.1051/0004-6361:20034219}

\bibitem[{{Sugimura} {et~al.}(2023){Sugimura}, {Matsumoto}, {Hosokawa},
  {Hirano}, \& {Omukai}}]{Sugimura+23}
{Sugimura}, K., {Matsumoto}, T., {Hosokawa}, T., {Hirano}, S., \& {Omukai}, K.
  2023, \apj, 959, 17, \dodoi{10.3847/1538-4357/ad02fc}

\bibitem[{{Susa}(2013)}]{Susa_13}
{Susa}, H. 2013, \apj, 773, 185, \dodoi{10.1088/0004-637X/773/2/185}

\bibitem[{Susa(2019)}]{Susa19}
Susa, H. 2019, The Astrophysical Journal, 877, 99,
  \dodoi{10.3847/1538-4357/ab1b6f}

\bibitem[{{Turk} {et~al.}(2011){Turk}, {Clark}, {Glover}, {Greif}, {Abel},
  {Klessen}, \& {Bromm}}]{Turk_2011}
{Turk}, M.~J., {Clark}, P., {Glover}, S.~C.~O., {et~al.} 2011, \apj, 726, 55,
  \dodoi{10.1088/0004-637X/726/1/55}

\bibitem[{{Vanzella} {et~al.}(2023){Vanzella}, {Loiacono}, {Bergamini},
  {Me{\v{s}}tri{\'c}}, {Castellano}, {Rosati}, {Meneghetti}, {Grillo},
  {Calura}, {Mignoli}, {Brada{\v{c}}}, {Adamo}, {Rihtar{\v{s}}i{\v{c}}},
  {Dickinson}, {Gronke}, {Zanella}, {Annibali}, {Willott}, {Messa}, {Sani},
  {Acebron}, {Bolamperti}, {Comastri}, {Gilli}, {Caputi}, {Ricotti},
  {Gruppioni}, {Ravindranath}, {Mercurio}, {Strait}, {Martis}, {Pascale},
  {Caminha}, {Annunziatella}, \& {Nonino}}]{vanzella+23}
{Vanzella}, E., {Loiacono}, F., {Bergamini}, P., {et~al.} 2023, \aap, 678,
  A173, \dodoi{10.1051/0004-6361/202346981}

\bibitem[{{Venditti} {et~al.}(2024){Venditti}, {Bromm}, {Finkelstein},
  {Graziani}, \& {Schneider}}]{Venditti+24}
{Venditti}, A., {Bromm}, V., {Finkelstein}, S.~L., {Graziani}, L., \&
  {Schneider}, R. 2024, \mnras, 527, 5102, \dodoi{10.1093/mnras/stad3513}

\bibitem[{Venditti {et~al.}(2023)Venditti, Graziani, Schneider, Pentericci,
  Di~Cesare, Maio, \& Omukai}]{Venditti+23}
Venditti, A., Graziani, L., Schneider, R., {et~al.} 2023, Monthly Notices of
  the Royal Astronomical Society, 522, 3809, \dodoi{10.1093/mnras/stad1201}

\bibitem[{{Vergara} {et~al.}(2021){Vergara}, {Schleicher}, {Boekholt},
  {Reinoso}, {Fellhauer}, {Klessen}, \& {Leigh}}]{Vergara21}
{Vergara}, M.~Z.~C., {Schleicher}, D.~R.~G., {Boekholt}, T.~C.~N., {et~al.}
  2021, \aap, 649, A160, \dodoi{10.1051/0004-6361/202140298}

\bibitem[{{Wang} {et~al.}(2020){Wang}, {Bose}, {Frenk}, {Gao}, {Jenkins},
  {Springel}, \& {White}}]{Wang+20}
{Wang}, J., {Bose}, S., {Frenk}, C.~S., {et~al.} 2020, \nat, 585, 39,
  \dodoi{10.1038/s41586-020-2642-9}

\bibitem[{{Welch} {et~al.}(2022){Welch}, {Coe}, {Diego}, {Zitrin},
  {Zackrisson}, {Dimauro}, {Jim{\'e}nez-Teja}, {Kelly}, {Mahler}, {Oguri},
  {Timmes}, {Windhorst}, {Florian}, {de Mink}, {Avila}, {Anderson}, {Bradley},
  {Sharon}, {Vikaeus}, {McCandliss}, {Brada{\v{c}}}, {Rigby}, {Frye}, {Toft},
  {Strait}, {Trenti}, {Sharma}, {Andrade-Santos}, \& {Broadhurst}}]{Welch+22}
{Welch}, B., {Coe}, D., {Diego}, J.~M., {et~al.} 2022, \nat, 603, 815,
  \dodoi{10.1038/s41586-022-04449-y}

\bibitem[{Wise(2019)}]{Wise_19}
Wise, J.~H. 2019, Contemporary Physics, 60, 145,
  \dodoi{10.1080/00107514.2019.1631548}

\bibitem[{{Wolfire} \& {Cassinelli}(1987)}]{Wolfire+87}
{Wolfire}, M.~G., \& {Cassinelli}, J.~P. 1987, \apj, 319, 850,
  \dodoi{10.1086/165503}

\bibitem[{{Wollenberg} {et~al.}(2020){Wollenberg}, {Glover}, {Clark}, \&
  {Klessen}}]{Wollenberg_20}
{Wollenberg}, K. M.~J., {Glover}, S. C.~O., {Clark}, P.~C., \& {Klessen}, R.~S.
  2020, \mnras, 494, 1871, \dodoi{10.1093/mnras/staa289}

\bibitem[{{Woods} {et~al.}(2017){Woods}, {Heger}, {Whalen}, {Haemmerl{\'e}}, \&
  {Klessen}}]{Woods_17}
{Woods}, T.~E., {Heger}, A., {Whalen}, D.~J., {Haemmerl{\'e}}, L., \&
  {Klessen}, R.~S. 2017, \apjl, 842, L6, \dodoi{10.3847/2041-8213/aa7412}

\bibitem[{Yajima {et~al.}(2023)Yajima, Abe, Fukushima, Ono, Harikane, Ouchi,
  Hashimoto, \& Khochfar}]{Yajima+23}
Yajima, H., Abe, M., Fukushima, H., {et~al.} 2023, Monthly Notices of the Royal
  Astronomical Society, 525, 4832, \dodoi{10.1093/mnras/stad2497}

\bibitem[{{Yong} {et~al.}(2013){Yong}, {Norris}, {Bessell}, {Christlieb},
  {Asplund}, {Beers}, {Barklem}, {Frebel}, \& {Ryan}}]{David+13}
{Yong}, D., {Norris}, J.~E., {Bessell}, M.~S., {et~al.} 2013, \apj, 762, 26,
  \dodoi{10.1088/0004-637X/762/1/26}

\bibitem[{{Yoshida} {et~al.}(2008){Yoshida}, {Omukai}, \&
  {Hernquist}}]{Yoshida_08}
{Yoshida}, N., {Omukai}, K., \& {Hernquist}, L. 2008, Science, 321, 669,
  \dodoi{10.1126/science.1160259}

\bibitem[{Yoshida {et~al.}(2006)Yoshida, Omukai, Hernquist, \&
  Abel}]{Yoshida_2006}
Yoshida, N., Omukai, K., Hernquist, L., \& Abel, T. 2006, The Astrophysical
  Journal, 652, 6, \dodoi{10.1086/507978}

\bibitem[{Yoshii {et~al.}(2022)Yoshii, Sameshima, Tsujimoto, Shigeyama, Beers,
  \& Peterson}]{Yoshii_2022}
Yoshii, Y., Sameshima, H., Tsujimoto, T., {et~al.} 2022, The Astrophysical
  Journal, 937, 61, \dodoi{10.3847/1538-4357/ac8163}

\bibitem[{Ása Skúladóttir {et~al.}(2021)Ása Skúladóttir, Salvadori,
  Amarsi, Tolstoy, Irwin, Hill, Jablonka, Battaglia, Starkenburg, Massari,
  Helmi, \& Posti}]{Skuladottir_2021}
Ása Skúladóttir, Salvadori, S., Amarsi, A.~M., {et~al.} 2021, The
  Astrophysical Journal Letters, 915, L30, \dodoi{10.3847/2041-8213/ac0dc2}

\end{thebibliography}

\end{document}